\documentclass[%
 reprint,
 superscriptaddress,
 amsmath,amssymb,
 aps,
 pra,         
 applied,     
  longbibliography
]{revtex4-2}

\usepackage{graphicx}
\usepackage{dcolumn}
\usepackage{bm}
\usepackage{xcolor}
\usepackage{booktabs}
\usepackage{tabularx}
\usepackage{multirow}
\usepackage{soul}
\usepackage{hyperref}
\usepackage{upgreek}
\usepackage[normalem]{ulem}

\nocite{apsrev41Control}

\begin{document}

\preprint{APS/123-QED}
\title{Towards Low-Energy Electron High-Resolution Spectroscopy\\ with Transition-Edge Sensors}

\author{R.~Ammendola}
\affiliation{INFN Sezione di Roma Tor Vergata, Roma, Italy}

\author{A.~Apponi}
\affiliation{INFN Sezione di Roma Tre, Roma, Italy}

\author{G.~Benato}
\affiliation{INFN Laboratori Nazionali del Gran Sasso (LNGS), L'Aquila, Italy}
\affiliation{Gran Sasso Science Institute (GSSI), L'Aquila, Italy}

\author{M.G.~Betti}
\affiliation{INFN Sezione di Roma, Roma, Italy}
\affiliation{Dipart. di Fisica, Sapienza Università di Roma, Roma, Italy}

\author{R.~Biondi}
\affiliation{INFN Laboratori Nazionali del Gran Sasso (LNGS), L'Aquila, Italy}
\affiliation{Gran Sasso Science Institute (GSSI), L'Aquila, Italy}

\author{P.~Bos}
\affiliation{Nationaal instituut voor subatomaire fysica (NIKHEF), Amsterdam, The Netherlands}
\affiliation{Dept. of Physics, University of Amsterdam, Amsterdam, The Netherlands}

\author{M.~Cadeddu}
\affiliation{INFN Sezione di Cagliari, Cagliari, Italy}

\author{A.~Casale}
\affiliation{Dept. of Physics, Columbia University, New York, USA}

\author{O.~Castellano}
\affiliation{INFN Sezione di Roma Tre, Roma, Italy}
\affiliation{Dipart. di Scienze, Università degli Studi di Roma Tre, Roma, Italy}

\author{G.~Cavoto}
\affiliation{INFN Sezione di Roma, Roma, Italy}
\affiliation{Dipart. di Fisica, Sapienza Università di Roma, Roma, Italy}

\author{L.~Cecchini}
\affiliation{INFN Sezione di Roma, Roma, Italy}

\author{E.~Celasco}
\affiliation{INFN Sezione di Genova, Genova, Italy}
\affiliation{Dipart. di Fisica, Università di Genova, Genova, Italy}

\author{M.~Chirico}
\affiliation{INFN Sezione di Roma, Roma, Italy}
\affiliation{Dipart. di Fisica, Sapienza Università di Roma, Roma, Italy}

\author{W.~Chung}
\affiliation{Princeton University, Princeton NJ, USA}

\author{A.G.~Cocco}
\affiliation{INFN Laboratori Nazionali del Gran Sasso (LNGS), L'Aquila, Italy}

\author{A.P.~Colijn}
\affiliation{Nationaal instituut voor subatomaire fysica (NIKHEF), Amsterdam, The Netherlands}
\affiliation{Dept. of Physics, University of Amsterdam, Amsterdam, The Netherlands}

\author{B.~Corcione}
\thanks{Corresponding author. E-mail: benedetta.corcione@uniroma1.it} 
\affiliation{INFN Sezione di Roma, Roma, Italy}
\affiliation{Dipart. di Fisica, Sapienza Università di Roma, Roma, Italy}

\author{N.~D'Ambrosio}
\affiliation{INFN Laboratori Nazionali del Gran Sasso (LNGS), L'Aquila, Italy}

\author{M.~D'Incecco}
\affiliation{INFN Laboratori Nazionali del Gran Sasso (LNGS), L'Aquila, Italy}

\author{G.~De~Bellis}
\affiliation{INFN Sezione di Roma, Roma, Italy}
\affiliation{Dipart. di Ingegneria Elettrica ed Energetica (DIEE), Sapienza Università di Roma, Roma, Italy}

\author{M.~De~Deo}
\affiliation{INFN Laboratori Nazionali del Gran Sasso (LNGS), L'Aquila, Italy}

\author{N.~de~Groot}
\affiliation{Dept. of Physics, Radboud University, Nijmegen, The Netherlands}

\author{A.~Esposito}
\affiliation{INFN Sezione di Roma, Roma, Italy}
\affiliation{Dipart. di Fisica, Sapienza Università di Roma, Roma, Italy}

\author{M.~Farino}
\affiliation{Princeton University, Princeton NJ, USA}

\author{S.~Farinon}
\affiliation{INFN Sezione di Genova, Genova, Italy}

\author{A.D.~Ferella}
\affiliation{INFN Laboratori Nazionali del Gran Sasso (LNGS), L'Aquila, Italy}
\affiliation{Dipart. di Fisica, Università degli Studi dell’Aquila, L’Aquila, Italy}

\author{L.~Ferro}
\affiliation{INFN Sezione di Cagliari, Cagliari, Italy}
\affiliation{Dipart. di Fisica, Università di Cagliari, Cagliari, Italy}

\author{L.~Ficcadenti}
\affiliation{INFN Sezione di Roma, Roma, Italy}

\author{G.~Galbato~Muscio}
\affiliation{INFN Sezione di Roma, Roma, Italy}
\affiliation{Dipart. di Ingegneria Astronautica, Elettrica ed Energetica, Sapienza Università di Roma, Roma, Italy}

\author{S.~Gariazzo}
\affiliation{INFN Sezione di Torino, Torino, Italy}
\affiliation{Dipart. di Fisica, Università di Torino, Torino, Italy}
\affiliation{Instituto de Fisica Corpuscular (IFIC - CSIC/UV), Paterna (Valencia), Spain}

\author{H.~Garrone}
\affiliation{INFN Sezione di Torino, Torino, Italy}
\affiliation{Istituto Nazionale di Ricerca Metrologica (INRiM), Torino, Italy}
\affiliation{Dipart. di Elettronica \& Telecomunicazioni (POLITO-ELN), Politecnico di Torino, Torino, Italy}

\author{F.~Gatti}
\affiliation{INFN Sezione di Genova, Genova, Italy}
\affiliation{Dipart. di Fisica, Università di Genova, Genova, Italy}

\author{F.~Malnati}
\affiliation{INFN Sezione di Torino, Torino, Italy}
\affiliation{Istituto Nazionale di Ricerca Metrologica (INRiM), Torino, Italy}
\affiliation{Dipart. Scienza Applicata e Tecnologia, Politecnico di Torino, Torino, Italy}

\author{G.~Mangano}
\affiliation{INFN Sezione di Napoli, Napoli, Italy}
\affiliation{Dipart. di Fisica, Università degli Studi di Napoli Federico II, Napoli, Italy}

\author{L.E.~Marcucci}
\affiliation{INFN Sezione di Pisa, Pisa, Italy}
\affiliation{Dipart. di Fisica, Università di Pisa, Pisa, Italy}

\author{C.~Mariani}
\affiliation{INFN Sezione di Roma, Roma, Italy}
\affiliation{Dipart. di Fisica, Sapienza Università di Roma, Roma, Italy}

\author{J.~Mead}
\affiliation{Nationaal instituut voor subatomaire fysica (NIKHEF), Amsterdam, The Netherlands}
\affiliation{Dept. of Physics, University of Amsterdam, Amsterdam, The Netherlands}

\author{G.~Menichetti}
\affiliation{Dipart. di Fisica, Università di Pisa, Pisa, Italy}

\author{M.~Messina}
\affiliation{INFN Laboratori Nazionali del Gran Sasso (LNGS), L'Aquila, Italy}

\author{E.~Monticone}
\affiliation{INFN Sezione di Torino, Torino, Italy}
\affiliation{Istituto Nazionale di Ricerca Metrologica (INRiM), Torino, Italy}

\author{M.~Naafs}
\affiliation{Nationaal instituut voor subatomaire fysica (NIKHEF), Amsterdam, The Netherlands}

\author{S.~Nagorny}
\affiliation{INFN Laboratori Nazionali del Gran Sasso (LNGS), L'Aquila, Italy}
\affiliation{Gran Sasso Science Institute (GSSI), L'Aquila, Italy}

\author{V.~Narcisi}
\affiliation{INFN Sezione di Roma Tor Vergata, Roma, Italy}
\affiliation{Dept. of Fusion and Technology for Nuclear Safety and Security, ENEA, Frascati, Italy}

\author{F.~Pandolfi}
\affiliation{INFN Sezione di Roma, Roma, Italy}

\author{R.~Pavarani}
\affiliation{INFN Sezione di Cagliari, Cagliari, Italy}
\affiliation{Dipart. di Fisica, Università di Cagliari, Cagliari, Italy}

\author{C.~Pepe}
\affiliation{Istituto Nazionale di Ricerca Metrologica (INRiM), Torino, Italy}
\affiliation{Instituto de Microelectrónica de Barcelona (IMB-CNM-CSIC), Barcelona, Spain}

\author{C.~P\'erez~de~los~Heros}
\affiliation{Dept. of Physics and Astronomy, Uppsala University, Uppsala, Sweden}

\author{O.~Pisanti}
\affiliation{INFN Sezione di Napoli, Napoli, Italy}
\affiliation{Dipart. di Fisica, Università degli Studi di Napoli Federico II, Napoli, Italy}

\author{F.M.~Pofi}
\affiliation{INFN Laboratori Nazionali del Gran Sasso (LNGS), L'Aquila, Italy}
\affiliation{Gran Sasso Science Institute (GSSI), L'Aquila, Italy}

\author{A.D.~Polosa}
\affiliation{INFN Sezione di Roma, Roma, Italy}
\affiliation{Dipart. di Fisica, Sapienza Università di Roma, Roma, Italy}

\author{I.~Rago}
\affiliation{INFN Sezione di Roma, Roma, Italy}

\author{M.~Rajteri}
\affiliation{INFN Sezione di Torino, Torino, Italy}
\affiliation{Istituto Nazionale di Ricerca Metrologica (INRiM), Torino, Italy}

\author{S.~Ritarossi}
\affiliation{INFN Sezione di Roma Tre, Roma, Italy}
\affiliation{Dipart. di Scienze, Università degli Studi di Roma Tre, Roma, Italy}

\author{N.~Rossi}
\affiliation{INFN Laboratori Nazionali del Gran Sasso (LNGS), L'Aquila, Italy}

\author{A.~Ruocco}
\affiliation{INFN Sezione di Roma Tre, Roma, Italy}
\affiliation{Dipart. di Scienze, Università degli Studi di Roma Tre, Roma, Italy}

\author{G.~Salina}
\affiliation{INFN Sezione di Roma Tor Vergata, Roma, Italy}

\author{A.~Santucci}
\affiliation{INFN Sezione di Cagliari, Cagliari, Italy}
\affiliation{Dipart. di Fisica, Università di Cagliari, Cagliari, Italy}

\author{M.~Sestu}
\affiliation{INFN Sezione di Cagliari, Cagliari, Italy}
\affiliation{Dipart. di Fisica, Università di Cagliari, Cagliari, Italy}

\author{A.~Tan}
\affiliation{Princeton University, Princeton NJ, USA}

\author{V.~Tozzini}
\affiliation{INFN Sezione di Pisa, Pisa, Italy}
\affiliation{CNR-Instituto Nanoscienze, Pisa, Italy}

\author{C.G.~Tully}
\affiliation{Princeton University, Princeton NJ, USA}

\author{I.~van~Rens}
\affiliation{Dept. of Physics, Radboud University, Nijmegen, The Netherlands}

\author{F.~Virzi}
\affiliation{INFN Laboratori Nazionali del Gran Sasso (LNGS), L'Aquila, Italy}
\affiliation{Dipart. di Fisica, Università degli Studi dell’Aquila, L’Aquila, Italy}

\author{G.~Visser}
\affiliation{Nationaal instituut voor subatomaire fysica (NIKHEF), Amsterdam, The Netherlands}

\author{M.~Viviani}
\affiliation{INFN Sezione di Pisa, Pisa, Italy}

\collaboration{PTOLEMY Collaboration}

\begin{abstract}
We present a study of the energy resolution of transition-edge sensors (TESs) for the detection of electrons in the 100~eV kinetic energy range. The TES is a Ti-Au bilayer with an active area of ($60 \times 60$) $\upmu$m$^2$ and a critical temperature of $\sim 80$~mK. The electron source is based on vertically-aligned multiwall carbon nanotubes located inside the cryostat, with electrons generated via field emission. For electrons in the $(92 - 99)$~eV kinetic energy range, we obtain a Gaussian energy resolution for fully-absorbed electrons of \textcolor{black}{($0.48 \pm 0.04 \pm 0.06$)~eV, where the first uncertainty is statistical and the second systematic}. When considering the full-width at half-maximum of the peak \textcolor{black}{of the signal amplitude distribution}, the corresponding resolution is of \textcolor{black}{($1.4 \pm 0.2 \pm 0.3$)~eV}. The former represents an improvement of $(\textcolor{black}{47}-60)\%$ with respect to previous results, and is mainly attributed to the reduction in the TES active area. The latter is instead an improvement of over a factor of 21, and is mainly due to the reduction in the emitting area of the electron source, which significantly suppresses electron back-scattering in proximity of the TES. These results represent a major milestone towards high-precision spectroscopy on low-energy electrons, which is a key objective for the PTOLEMY experiment. 
\end{abstract}

\maketitle

\section{\label{sec:introduction}Introduction}

\noindent Transition-Edge Sensors (TESs) are microcalorimeters with intrinsic energy resolution capable of detecting single particles over a wide energy spectrum \cite{intro1, intro2, intro3, intro4, cresst, cdms, holmes}. They consist of a superconductive thin film operated at the steep superconductive transition near its critical temperature $T_\text{C}$, where the resistance of the material is most sensitive to temperature changes \cite{irwin}. When a particle hits the TES, its energy is deposited into the device via excitations of the lattice and the electronic system, leading to an increase in the TES temperature. This heating produces measurable changes in the TES resistance, which are proportional to the absorbed energy. Such detection mechanism provides the TES with an intrinsic energy resolution, making it particularly suitable to detect small energy transfers. TES devices have been extensively used as high-precision low-energy single-photon detectors, on which they have reached a Gaussian energy resolution below 50 meV for 0.8 eV photons \cite{fotonires, fotonires2}. \\

Recently TES devices have been tested for electron detection, as they offer unique possibilities compared to other devices such as micro-channel plates or silicon-based detectors. In fact TES devices, in principle, have no dead layers and a unitary fill factor \cite{mcp1, mcp2, sdd}, enabling full collection and measurement of the energy deposited in the device. Moreover, the dark count \textcolor{black}{rate} of such devices is $\mathcal{O}$(mHz) \cite{dark_counts}, making them suitable to search for rare events; \textcolor{black}{they can achieve timing resolutions of a few nanoseconds \cite{timeres} and a maximum count rate of $\mathcal{O}$(MHz) \cite{maxcount}}. These characteristics are crucial to make TES devices excellent detectors with intrinsic energy resolution, which is why this technology could play a pivotal role in low-energy electron spectroscopy \cite{patel}, microcalorimetry \cite{microcal1, microcal2} and experiments studying $\beta$-decay and neutrino physics \textcolor{black}{\cite{holmes}}. Among the latter, the PTOLEMY project \cite{ptolemy1, ptolemy2, ptolemy3} plans to \textcolor{black}{measure the mass of the neutrino and detect the cosmic neutrino background} by analyzing the endpoint of the $\beta^-$ decay spectrum of atomic tritium. To reach these objectives, \textcolor{black}{the TES developed for the PTOLEMY experiment should achieve} a Gaussian energy resolution of 50 meV on electrons with a kinetic energy of 10~eV. To date, a Gaussian energy resolution $\sigma_\text{e} > 17$~eV has been achieved on electrons in the $300 - 2000$~eV energy range \cite{patel}, and $0.7 < \sigma_\text{e} <1.6$~eV on electrons with kinetic energy $91 \leq E_\text{e} \leq 101$~eV \cite{articolo, ncc}. These latter results appear promising for the PTOLEMY application, so modifications to the experimental setup were implemented to investigate them further. \\

This work presents new results in which a TES device is used to measure the energy of electrons with kinetic energy in the $\sim 100$~eV range. Compared to the results presented in the previous work of \cite{articolo}, where a TES with an active area of $(100 \times 100)$~$\upmu$m$^2$ and an electron source based on field emission from a 9~mm$^2$ carbon nanotube sample were employed, this work uses a $(60 \times 60)$~$\upmu$m$^2$ TES  and a 1~mm$^2$ carbon nanotube source. As shown below, the smaller TES improves the intrinsic energy resolution for fully absorbed electrons, while the reduced source size decreases electron backscattering in the proximity of the TES, thereby greatly suppressing signals from electrons with sub-nominal energy. Together, these changes reduce the width of the electron energy peak by more than an order of magnitude.

\section{\label{sec:experimental setup}Experimental Setup}
\noindent The measurements presented in this work are performed at the Innovative Cryogenic Laboratory of the Istituto Nazionale di Ricerca Metrologica (INRiM) in Torino (Italy).  An adiabatic demagnetization refrigerator cryostat is used to maintain the TES at a stable bath temperature. Since magnetic fields affect the trajectories of electrons and the read-out system based on dc-SQUIDs \cite{squid}, a magnetic shield \textcolor{black}{made of Amumetal 4K} is mounted within the cryostat to surround the whole working area, \textcolor{black}{which consists of the 30 mK plate hosting both the TES experimental setup and the SQUID read-out system. The shield can suppress the Earth magnetic field by a factor $\sim$ 60.} \\

The TES used for this work is a bilayer made of 15~nm of titanium and  30~nm of gold, with an active area of $(60 \times 60)$ $\upmu$m$^2$ and a critical temperature $T_C$ of $\sim$ 80~mK.  More details on its fabrication are reported in~\cite{articolo}. Its dark count rate is \textcolor{black}{$\lesssim$ 0.8 mHz in the full energy range of the experiment of 92$-$99 eV, which is negligible relative to the signal rate at the selected trigger threshold, as expected from this kind of devices \cite{dark_counts}}. \\

A sample of vertically-aligned multiwall carbon nanotubes (CNTs) is used as an electron source, as they have proven to be a reliable field-emitter at cryogenic temperatures \cite{articolo_cnt}. They were synthesized in the INFN laboratory ‘TITAN’ at Sapienza University of Rome \cite{titan1, titan2, titan3, titan4, articolo_cnt} by chemical vapor deposition on a 500 $\upmu$m silicon substrate, following the procedure described in \cite{articolo_cnt}, and covering an area of $\sim$ 1 mm$^2$. \\

The CNTs are positioned in front of the TES and are voltage-biased with a negative voltage $V_\text{CNT}$ to emit electrons through field emission, while the TES is electrically grounded through the cryostat and thermally connected to it. As explained in more detail in \cite{articolo}, to adapt the design of the TES for electron detection, a \textcolor{black}{50 nm thick} gold shield is deposited everywhere on the substrate except on the active area of the device, in order to protect both the wiring and the insulating substrate from impinging electrons, as the active area of the TES is much smaller than that of the CNTs. \\

According to Fowler-Nordheim theory \cite{fn_theory}, the field-emitted current increases exponentially with the applied voltage. A large electron current hitting the gold shield layer has two adverse effects: first it causes an excessive increase in the TES temperature, which in turn decreases the TES dynamic range; secondly, some electrons are back-scattered from the gold shield layer and then enter the TES with only a fraction of their nominal energy, thus broadening their otherwise monochromatic energy spectrum. Both these aspects can be mitigated by decreasing the area of the CNT source, as fewer electrons are sent towards the device.\\

The kinetic energy of field-emitted electrons when they reach the TES is derived in \cite{articolo} as:
\begin{equation}
E_\text{e} = eV_\text{CNT} - \phi_\text{TES}
\label{eq energy}
\end{equation}
where $e$ is the elementary electric charge and $\phi_\text{TES}$ is the TES work function. The latter was measured with ultra-violet photoemission spectroscopy in the LASEC lab of Roma Tre University \cite{articolo_cnt}, yielding a value of $\phi_\text{tes}$ = (4.38 $\pm$ 0.03) eV. Moreover, CNTs can be considered a monochromatic source, as the energy spread of the field-emitted electrons is negligible at the temperature and electric field ranges of the experiment \cite{monochromaticity}. \textcolor{black}{Since field emission has an exponential dependence on the electric field, below a minimum $V_\text{CNT}$ the rate of emitted electrons is too low to detect them. This defines from Eq. \ref{eq energy} a minimum $E_\text{e}$, which in our setup is $E_\text{e}^\text{min}\sim 92$~eV.} \\

The TES is \textcolor{black}{operated} at $35\%$ of its normal resistance \textcolor{black}{$R_\text{N} \sim 0.2~\Omega$}, where it was found to be optimal in terms of energy resolution. To ensure stable working conditions, the TES is operated in \textcolor{black}{negative} electrothermal feedback, \textcolor{black}{voltage-biased exploiting a shunt resistor of 20 m$\Omega$, and} the bath temperature is approximately set at half the critical temperature of the device. \textcolor{black}{In series with the TES there is a 6~nH inductance, coupled to the read-out which consists of an array of 16 dc-SQUIDs \cite{squid}. A schematic of the circuitry can be found in \cite{articolo}.}
To guarantee a more stable voltage supply a low-pass filter with cutoff frequency of 200~Hz is introduced on the power line of the CNTs at the 3~K stage of the cryostat\textcolor{black}{, which in turn ensures a more stable field-emission current}. Moreover, to reduce the high-frequency noise on the pulse shapes, the signals are pre-processed with a Moku:Lab FPGA-based unit by Liquid Instruments, using a Bessel 100~kHz low-pass filter. Since no such filter was used in the work presented in \cite{articolo}, we have re-processed those data by introducing an additional software-based RC low-pass filter with a cutoff frequency of 100~kHz, so as to allow for a more rigorous comparison.\\

\begin{figure}[t]
\includegraphics[width=0.92\columnwidth]{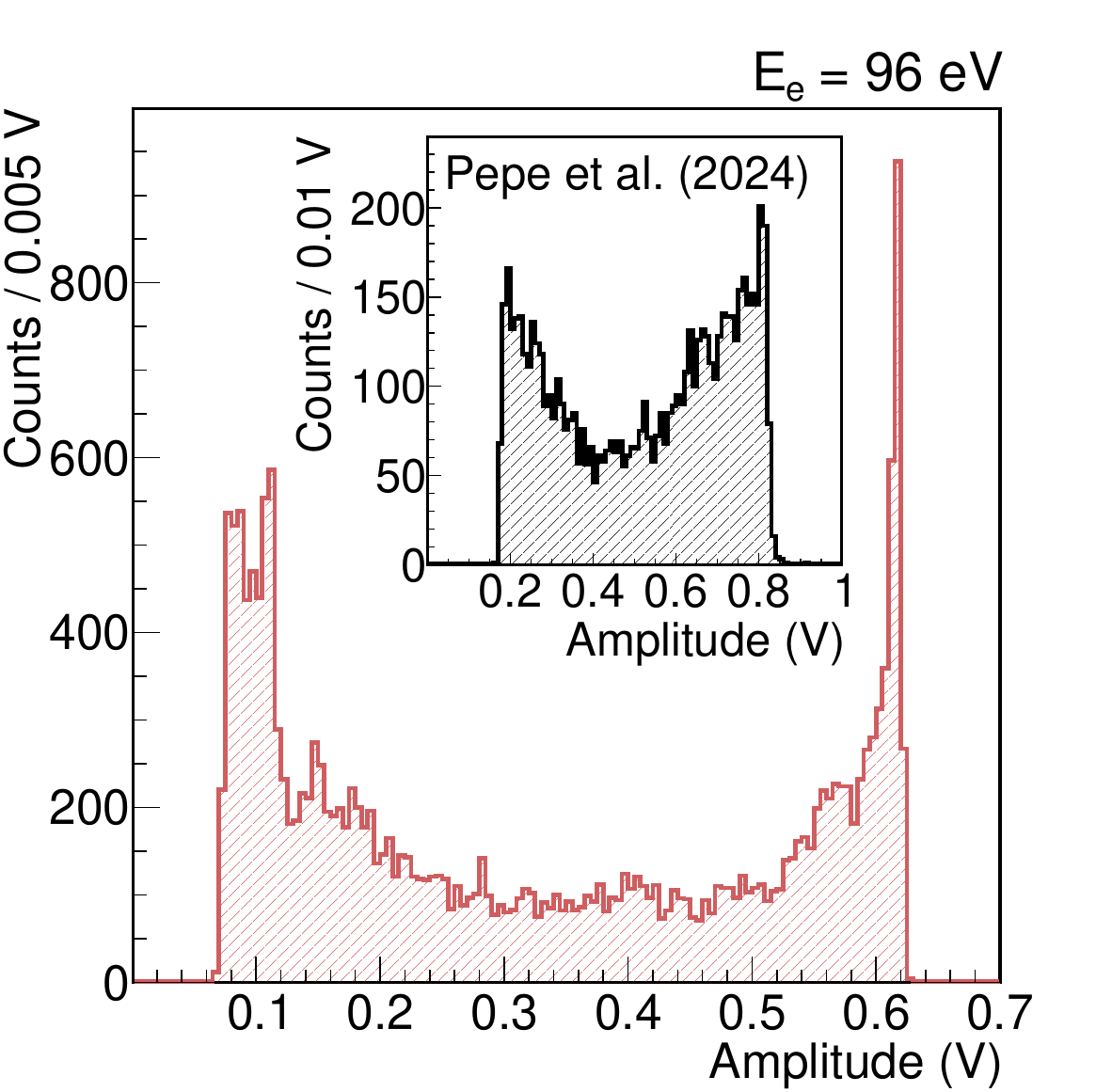}
\caption{Signal amplitude spectrum for 96-eV electrons (red histogram), compared to the same spectrum obtained in the previous measurement \cite{articolo}(black histogram in the inset). \textcolor{black}{The different amplitude scale in the two distributions  is due to the fact that the red distribution refers to low-pass filtered signals, while the black distribution refers to raw pulse shapes.}}
\label{fig spectrum_comparison}
\end{figure}

\begin{figure}[ht]
\includegraphics[width=0.9\columnwidth]{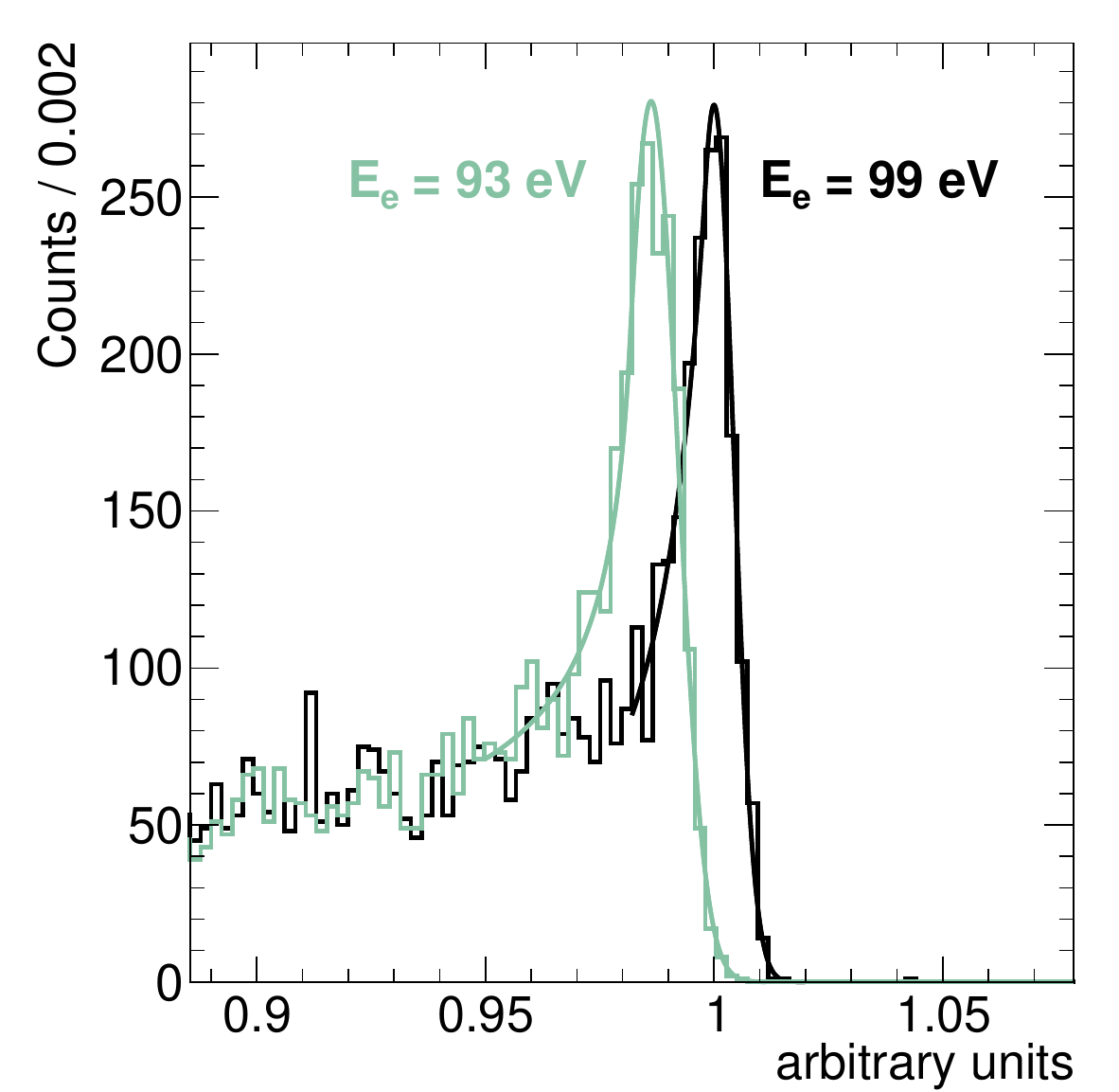}
\caption{Example fits of the electron absorption peaks: the green data corresponds to electrons with a kinetic energy $E_{\text{e}} = 93$~eV, the black data to $E_{\text{e}} = 99$~eV. \textcolor{black}{More details about the fit are given in the text.}}
\label{fig fit}
\end{figure}

\section{\label{sec:data analysis}Data Analysis}

\begin{figure}[tb]
\includegraphics[width=0.85\columnwidth]{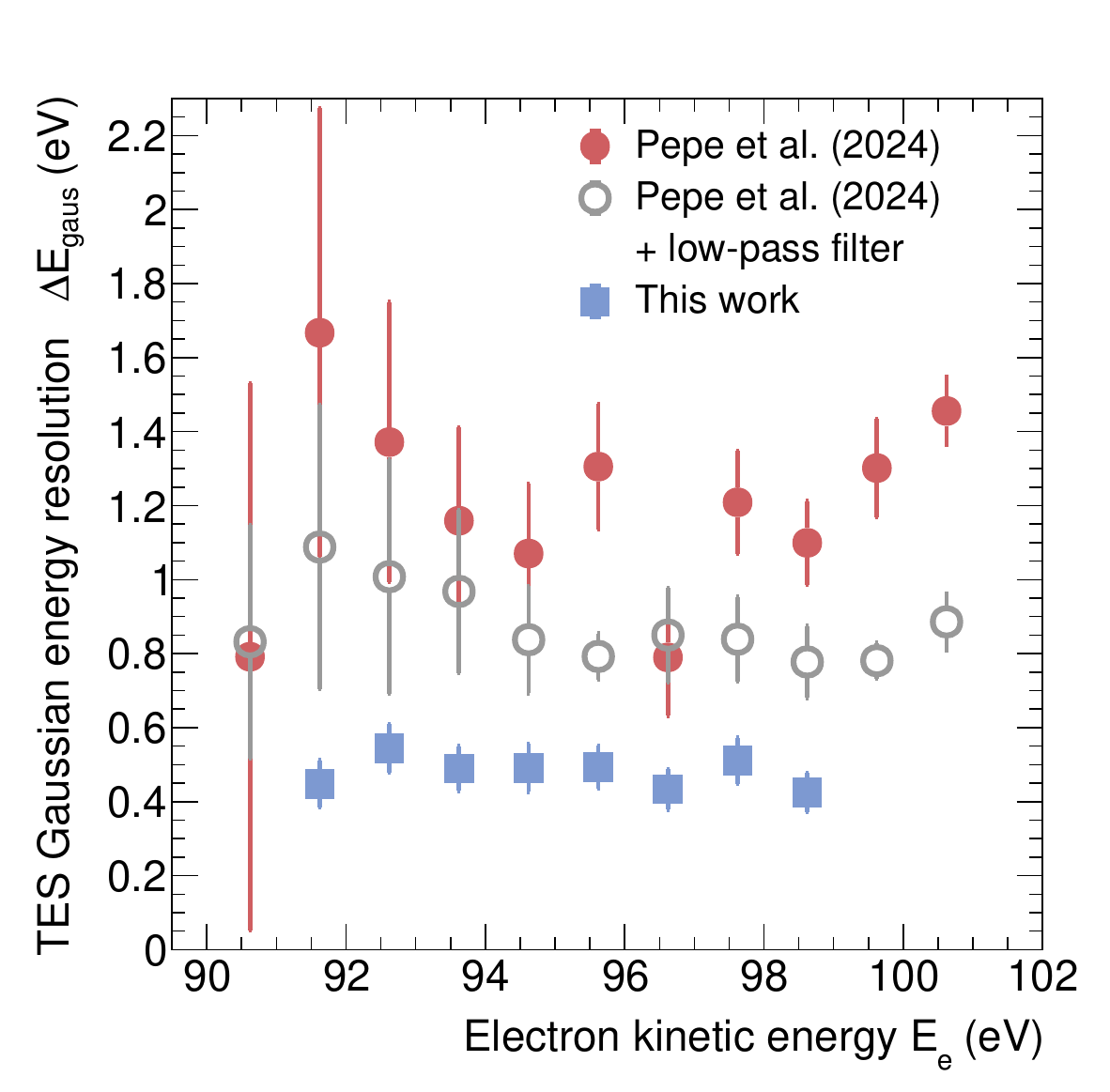}
\includegraphics[width=0.85\columnwidth]{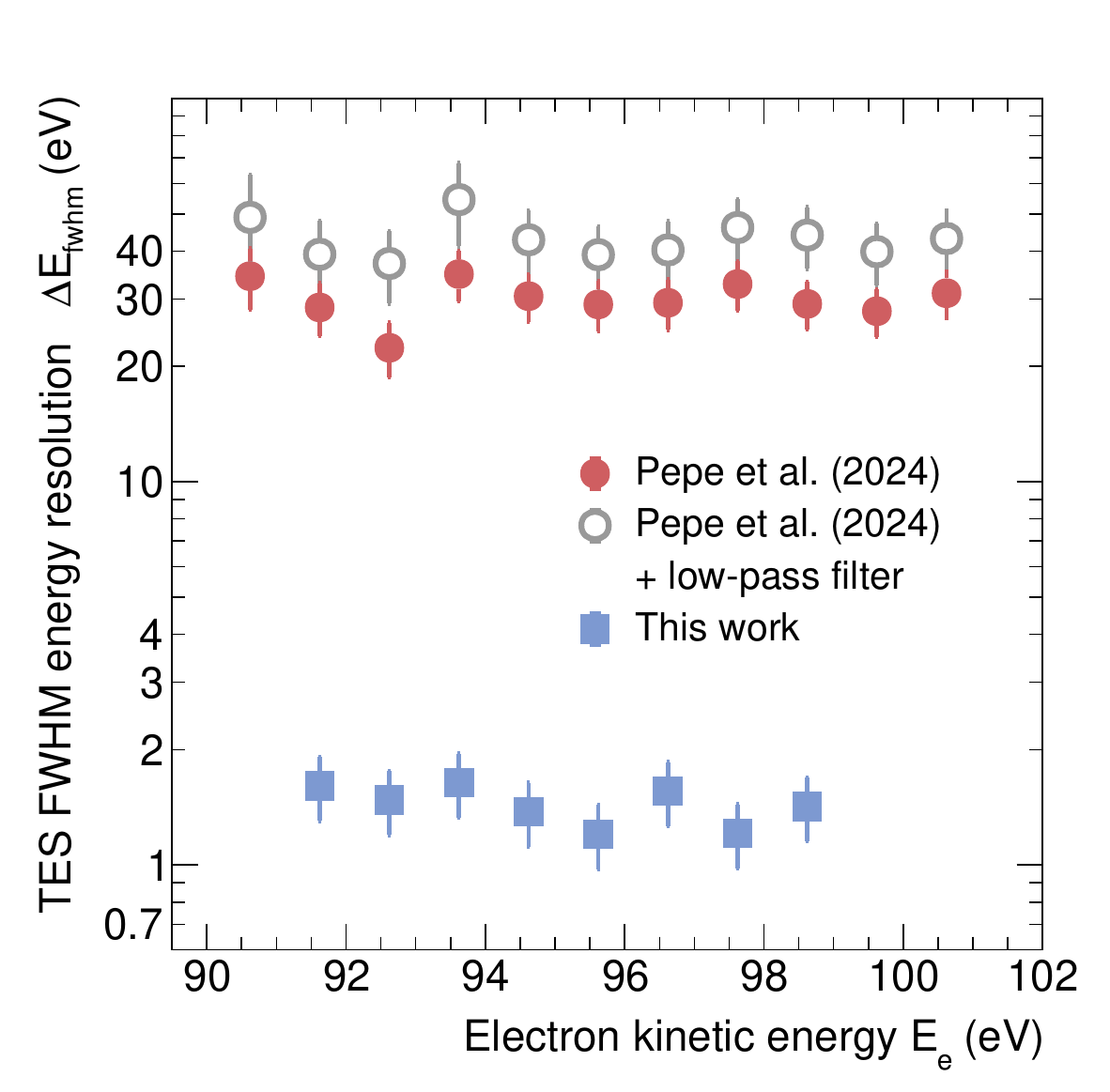}
\caption{Top: TES Gaussian energy resolution $\Delta E_\text{gaus}$ for electrons, as a function of their kinetic energy $E_{\text{e}}$. Bottom: TES full-width at half-maximum energy resolution $\Delta E_\text{fwhm}$ for electrons, as a function of their kinetic energy $E_{\text{e}}$\textcolor{black}{, with logarithmic scale on the y-axis.}}
\label{fig resol_sigma_fwhm}
\end{figure}

\noindent The red histogram in Figure \ref{fig spectrum_comparison} shows the TES signal amplitude spectrum for electrons with a kinetic energy of 96~eV, and is compared to the same spectrum obtained in~\cite{articolo}, which is shown in black in the inset. In both the distributions, the peak in the high-amplitude region is associated to events in which the electrons deposit all their energy inside the device, so it was \textcolor{black}{referred to as} “full-absorption peak". 
As can be seen, the tail to the left of the full-absorption peak is significantly less pronounced in the data of this work. This is mainly due to the smaller CNT sample size, which leads to fewer electrons back-scattering off the gold shield and hitting the TES active area with lower-than-nominal energy. \\

    The absorption peaks of the distributions are fitted with a Crystal Ball function \cite{cb}: its right side with respect to the peak position ($\upmu$) is  Gaussian with standard deviation $\sigma$; its left side is Gaussian up to a transition point, where it becomes a power-law tail. In particular, $\sigma$ \textcolor{black}{is the only parameter describing the right tail of the full-absorption peak, which is broadened by the energy resolution of the device and the energy spread of the electron source. As previously stated, the CNTs in our working conditions emit monochromatic electrons, so $\sigma$} is dominated by the energy resolution of the device. The power-law tail \textcolor{black}{instead} describes partially-absorbed electrons. \textcolor{black}{The full-absorption peak is fitted in a range between $\pm$ 2\% with respect to the peak position.}  Figure \ref{fig fit} shows two typical fits to the absorption peaks: the data corresponding to electrons with a kinetic energy of 93~(99)~eV is shown in green~(black). As for the data presented in \cite{articolo} and post-filtered, they are fitted with an asymmetric Gaussian function, as explained in the previous work.

\begin{table*}[tb]
\centering
\newcolumntype{Y}{>{\centering\arraybackslash}X}
\begin{tabularx}{0.9\textwidth}{Y Y Y Y Y Y}
\hline
\multirow{2}{*}{\textbf{Dataset}}& \textbf{TES area} & \textbf{CNT area} &
\textbf{$E_\text{e}$} & \textbf{$\langle\Delta E_\text{gaus}\rangle$} &
\textbf{$\langle\Delta E_\text{fwhm}\rangle$} \\
 & ($\upmu$m$^2$) & (mm$^2$) &
(eV) & (eV) & (eV)\\
\hline
Pepe et al. 2024 (raw) & $100 \times 100$ & 9 & $91 - 101$ & \textcolor{black}{$1.2 \pm 0.3$} & \textcolor{black}{$30 \pm 4$} \\
Pepe et al. 2024 (low-pass filter) & $100 \times 100$ &  9 & $91 - 101$ &  \textcolor{black}{$0.9 \pm 0.1$} & \textcolor{black}{$43 \pm 5$} \\ 
This work  &  $60 \times 60$ & 1 & $92 - 99$ & \textcolor{black}{$0.48 \pm 0.04$} & \textcolor{black}{$1.4 \pm 0.2$} \\
\hline
\end{tabularx}
\caption{Summary of TES energy resolution results. The values reported in the last two columns correspond, respectively, to the average $\Delta E_\text{gaus}$ and $\Delta E_\text{fwhm}$ values obtained on the studied electron energy range. Uncertainties are statistical in nature.}
\label{tabella-results}
\end{table*}

The electron energy resolution is defined using both $\sigma$ and the full-width at half-maximum FWHM, respectively as $\Delta E_\text{gaus}$ = ($\sigma$/$\upmu$) $\cdot$ $E_\text{e}$ and $\Delta E_\text{fwhm}$ = (FWHM/$\upmu$) $\cdot$ $E_\text{e}$, where the electron kinetic energy $E_\text{e}$ is defined in Equation \ref{eq energy}. \textcolor{black}{The FWHM is obtained directly from the histogram of signal amplitude.} Figure \ref{fig resol_sigma_fwhm} shows the energy resolutions $\Delta E_\text{gaus}$ and $\Delta E_\text{fwhm}$ achieved by this work~(blue squares). These are in turn compared to the data recorded in~\cite{articolo}, both in their original form~(full red circles) and after the application of a 100~kHz low-pass filter~(open gray circles). These results are also summarized in Table \ref{tabella-results}. 
\\

As discussed in the Experimental Setup section, the data reported in \cite{articolo} were re-processed using a low-pass filter matched to the signal conditioning employed in the present work, in order to enable a consistent comparison. Under these conditions, the re-processed dataset exhibits an improvement of approximately 26\% in $\Delta E_\text{gaus}$ with respect to the raw data. Using the present setup, we obtain a Gaussian energy resolution of \textcolor{black}{$\langle\Delta E_\text{gaus}\rangle=~(0.48 \pm 0.04 \pm 0.06$)~eV}, where the average has been made across the full energy range, and the first uncertainty is statistical and the second is systematic. This corresponds to a 60\% (46\%) improvement over the raw (filtered) data of our previous work. The improvement with respect to the filtered data is roughly in line with the expected $\sim$ 40\% improvement in TES energy resolution coming from the reduction in the TES active area. In fact, the intrinsic energy resolution of a TES is known to be proportional to the square root of its active area\textcolor{black}{: in fact, the intrinsic energy resolution of a TES is known to scale with the square root of its heat capacity, which in turn depends on the TES volume \cite{irwin}. Therefore, keeping the film thickness constant, reducing the TES active area lowers its heat capacity and results in a larger temperature rise for the same absorbed energy, ultimately improving the detector response and energy resolution.} These results indicate that the observed performance enhancement in the present work is primarily driven by detector design rather than by differences in signal filtering.

The FWHM of the absorption peak, on the other hand, is strongly influenced by the reduced size of the CNT emitting area. As shown in Figure~\ref{fig resol_sigma_fwhm} \textcolor{black}{bottom}, we obtain, when averaging across the full electron energy range, \textcolor{black}{$\langle\Delta E_\text{fwhm}\rangle = (1.4 \pm 0.2 \pm 0.3)$~eV}. This corresponds to an improvement of a factor 21 (\textcolor{black}{31}) with respect the raw (filtered) data of our previous work. It is worth noting that, differently from the case of $\Delta E_{\text{gaus}}$, in this case the resolution of our previous setup is worsened by the application of the low-pass filter. In fact, while the width of the right tail of the absorption peak is mainly driven by \textcolor{black}{statistical resolution effects} and therefore benefits from filtering, its left tail, in the case of \cite{articolo}, is largely dominated by unwanted signals \textcolor{black}{of partially-absorbed electrons} that persist even after filtering\textcolor{black}{, whose amplitude distribution is modified by the effect of the filter and happens to lead to a wider left tail, and consequently to a worse $\Delta E_\text{fwhm}$.}  \\

The main systematic uncertainty concerns the choice of the \textcolor{black}{fitting} model. To estimate this, the systematic uncertainty of the $\Delta E_\text{gaus}$ is calculated by evaluating the variation in energy resolution obtained by using an asymmetric Gaussian instead of a Crystal Ball. This results in a value of 12\% for the data of this work, determined as the difference in $\Delta E_\text{gaus}$ obtained with the two different functions, averaged across \textcolor{black}{the tested electron energies}. To estimate the systematic uncertainty on $\Delta E_\text{fwhm}$, \textcolor{black}{the FWHM derived directly from the distribution of the data is compared to the FWHM analitically calculated from the parameters of the fit.} The resulting systematic uncertainties are 19\%, 14\% and 18\% for the new data, the previous data, and the previous data post-filtered, respectively, all averaged on the \textcolor{black}{different applied voltages}. \textcolor{black}{Moreover, to validate all the results obtained fitting the histograms, the effect of different binnings and fitting ranges are studied. The discrepancy among different energy resolution results is $\sim$ 3\%, which is added in quadrature to the other sources of uncertainty.} The total uncertainty reported in the error bars of Figure \ref{fig resol_sigma_fwhm} represents the quadrature sum of the statistical contribution and the systematic contribution. The $\Delta E_\text{gaus}$  of the data already presented in \cite{articolo} (full red circles) is shown as originally published, and to allow for a more rigorous comparison, no  additional uncertainty is added to the corresponding filtered dataset (open gray circles), to which the same uncertainties reported in \cite{articolo} are applied. \\

\begin{figure}[t]
\includegraphics[width=0.9\columnwidth]{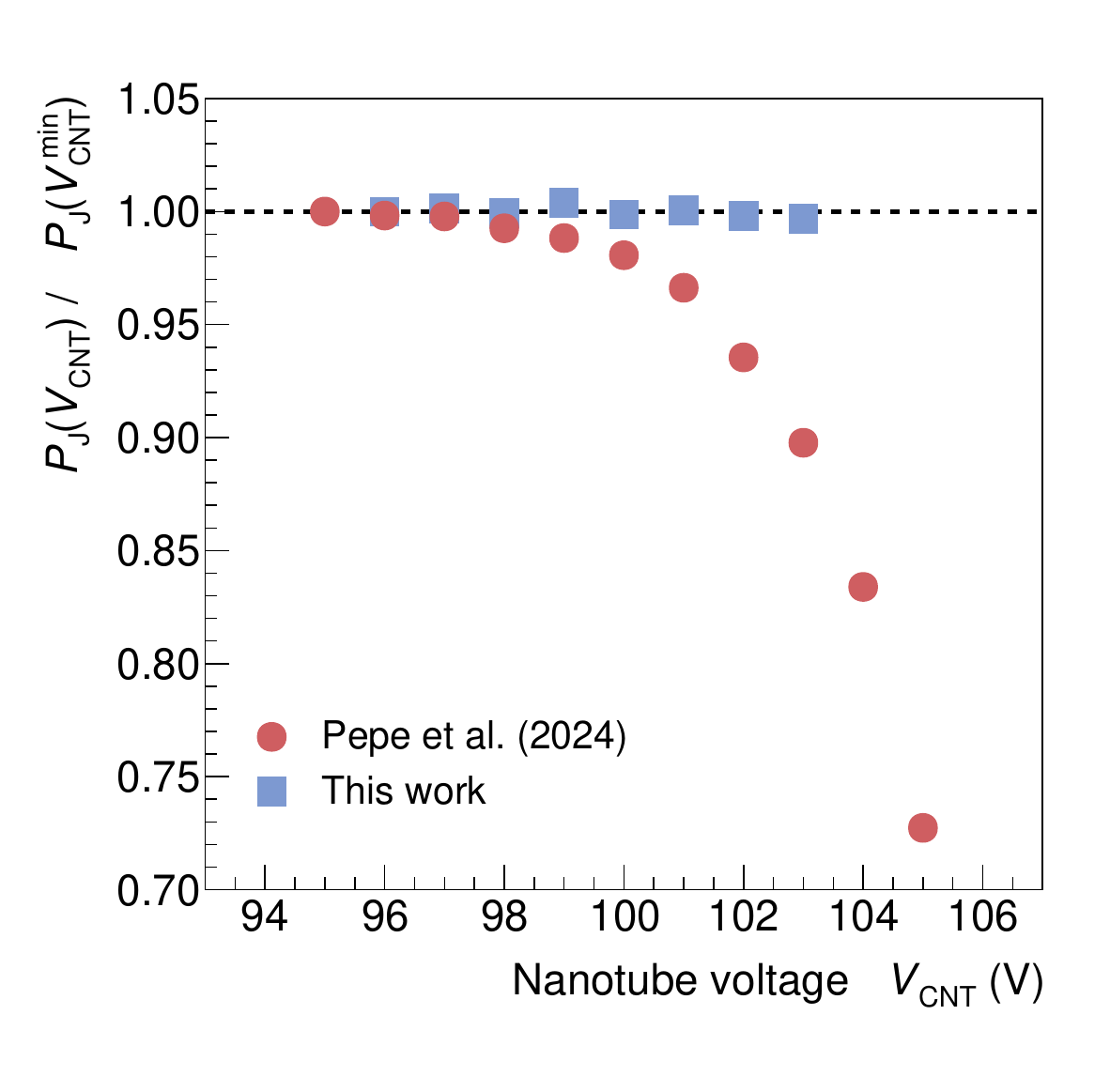}
\caption{Joule power $P_\text{J}$ needed to bring the TES to its working point for different values of $V_\text{CNT}$, normalized to the Joule power needed for the lowest $V_\text{CNT}$ value. A black dashed line corresponding to unity is plotted to guide the eye.}
\label{fig joulepower}
\end{figure}

\begin{figure}[ht!]
\includegraphics[width=0.9\columnwidth]{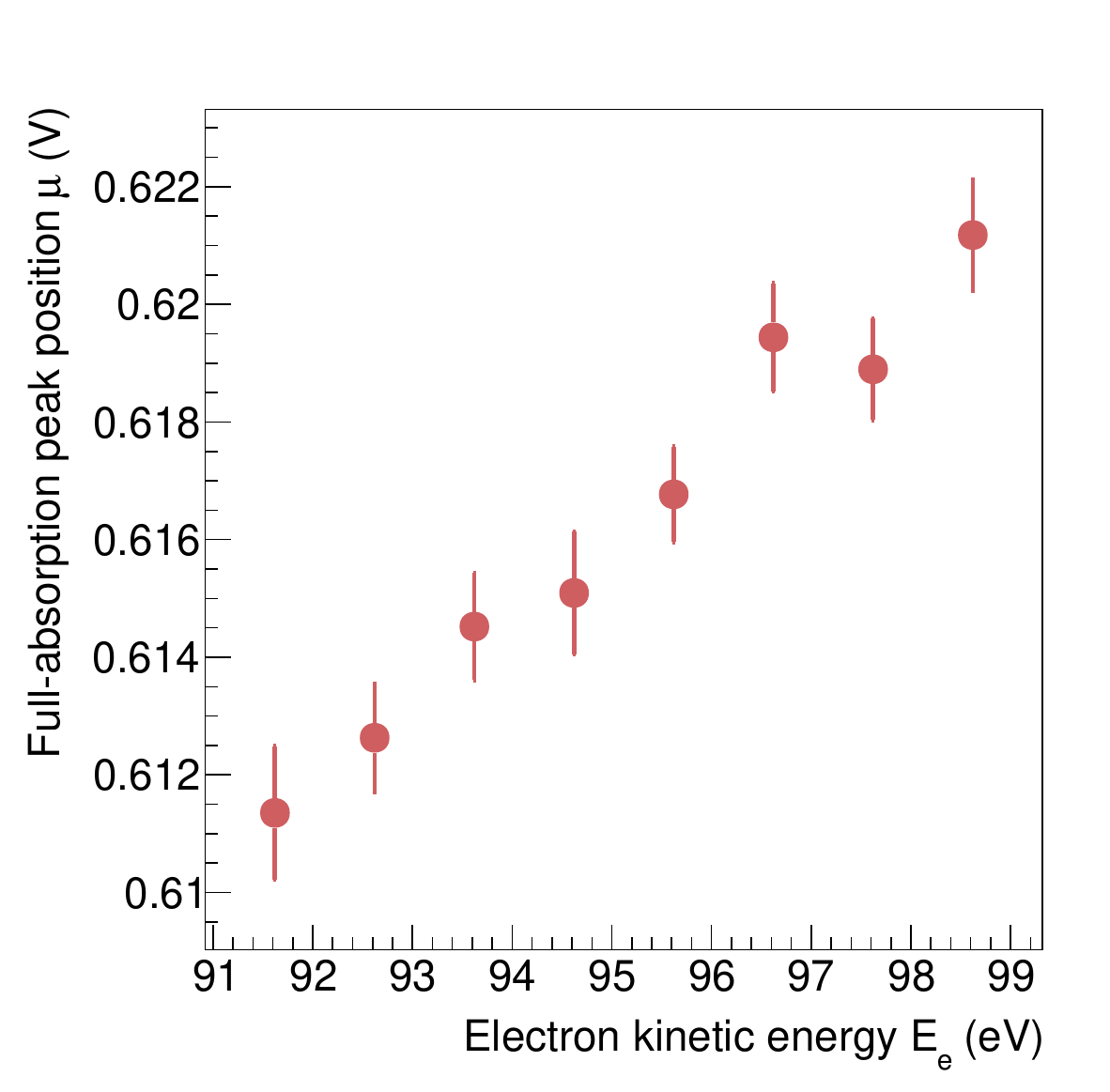}
\caption{\textcolor{black}{Full-absorption peak position $\upmu$ as a function of electron kinetic energy $E_e$.}}
\label{fig mu}
\end{figure}

In addition to improving the TES energy resolution, reducing the emitting surface of the electron source mitigates the heating effect associated with field-emission observed in \cite{articolo}. This aspect is illustrated in Figure~\ref{fig joulepower}, where the values of Joule power $P_\text{J}$ required to bring the TES to its working point on the superconductive transition are plotted with respect to the applied bias voltage $V_\text{CNT}$. In order to compare different measurements, each dataset is normalized to the Joule power dissipated at the lowest $V_\text{CNT}$ analyzed. As can be seen, in the data of this work (blue squares) no relevant heating is present, as the Joule power needed to polarize the TES at its operating temperature remains stable. In the previous data (red circular markers), instead, an exponential reduction of the Joule power $P_\text{J}$ dissipated through the TES with increasing  $V_\text{CNT}$ can be observed, indicating a corresponding increase of the substrate temperature. Operating the TES under consistent conditions ensures that measurements performed at different $V_\text{CNT}$ values, corresponding to different electron kinetic energies, can be rigorously compared without applying additional corrections. We have furthermore verified that the peak position $\upmu$ of the Crystal Ball fit function increases linearly with $V_\text{CNT}$, confirming the stability of the TES working conditions\textcolor{black}{, as can be appreciated in Figure~\ref{fig mu}.}

\section{\label{sec:conclusions}Conclusions}
\noindent We have presented  \textcolor{black}{a new experimental setup for the PTOLEMY experiment to investigate the energy resolution of a TES characterized with monochromatic electrons in the 100-eV kinetic energy range.} Compared to our previous work~\cite{articolo}, we have employed a TES with a smaller active area of $(60\times 60)$~$\upmu$m$^2$, and a smaller electron source, based on field emission from a $(1\times 1)$~mm$^{2}$ chip of vertically-aligned carbon nanotubes (CNTs). We report a remarkable improvement in the energy resolution, both in the Gaussian width of the full-absorption peak ($\Delta E_{\text{gaus}}$), and in the full-width at half-maximum broadness of the peak ($\Delta E_{\text{fwhm}}$). When averaging the results over the full energy range, we obtain \textcolor{black}{$\langle\Delta E_\text{gaus}\rangle = (0.48 \pm  0.04 \pm 0.06$)~eV}, which corresponds to a $60\%$ improvement with respect to the raw data of our previous work, and to a \textcolor{black}{$47\%$} improvement with respect to that same data after the application of a low-pass filter with a frequency of 100~kHz. This improvement is in line with the expectations deriving from the reduction of the TES active area. Taking the same average on $\Delta E_{\text{fwhm}}$ yields~\textcolor{black}{$\langle\Delta E_\text{fwhm}\rangle = (1.4 \pm  0.2 \pm 0.3$)~eV}, which corresponds to an improvement of a factor 21 (\textcolor{black}{31}) with respect to the raw~(filtered) data of our previous work. This large improvement, of over an order of magnitude, is mainly due to the reduction in the CNT source size, and represents a major step forward in the spectroscopy of low-energy electrons.\\

To keep testing the potential of TES devices with electrons, lower energies will be tested by inserting a decelerating plate inside the experimental setup. This will allow us to reach the energy range $E_{\text{e}} \sim 10$~eV required by the PTOLEMY collaboration, \textcolor{black}{further decreasing the size of the TES detectors to improve their intrinsic energy resolution. Since the devices will be smaller, a larger amount of them will be required to geometrically cover the full spatial distribution of electrons at the end of the PTOLEMY filter, resulting in an array of TESs. Consequently, the total rate of electrons originating from the tritium source will be distributed across the entire TES array rather than concentrated on an individual device.}
  
\section{\label{sec:acknowledgements}Acknowledgements}
\noindent The authors are grateful to Elio Bertacco, Martina Marzano, Matteo Fretto and Ivan De Carlo for technical support. This research was partially funded by the contribution of grant 62313 from the John Templeton Foundation, by PRIN grant ‘ANDROMeDa’ (PRIN 2020Y2JMP5) and PRIN grant ‘TES4e’ (PRIN 2022 202285H5MT) of Ministero dell’Università e della Ricerca, and from the EC project ATTRACT (Grant Agreement No. 777222). We also acknowledge funding from the Agencia Estatal de Investigacion (Ref. CNS2023-145234).

\bibliographystyle{apsrev4-2-titles}
\bibliography{references}

\begin{thebibliography}{34}%
\makeatletter
\providecommand \@ifxundefined [1]{%
 \@ifx{#1\undefined}
}%
\providecommand \@ifnum [1]{%
 \ifnum #1\expandafter \@firstoftwo
 \else \expandafter \@secondoftwo
 \fi
}%
\providecommand \@ifx [1]{%
 \ifx #1\expandafter \@firstoftwo
 \else \expandafter \@secondoftwo
 \fi
}%
\providecommand \natexlab [1]{#1}%
\providecommand \enquote  [1]{``#1''}%
\providecommand \bibnamefont  [1]{#1}%
\providecommand \bibfnamefont [1]{#1}%
\providecommand \citenamefont [1]{#1}%
\providecommand \href@noop [0]{\@secondoftwo}%
\providecommand \href [0]{\begingroup \@sanitize@url \@href}%
\providecommand \@href[1]{\@@startlink{#1}\@@href}%
\providecommand \@@href[1]{\endgroup#1\@@endlink}%
\providecommand \@sanitize@url [0]{\catcode `\\12\catcode `\$12\catcode
  `\&12\catcode `\#12\catcode `\^12\catcode `\_12\catcode `\%12\relax}%
\providecommand \@@startlink[1]{}%
\providecommand \@@endlink[0]{}%
\providecommand \url  [0]{\begingroup\@sanitize@url \@url }%
\providecommand \@url [1]{\endgroup\@href {#1}{\urlprefix }}%
\providecommand \urlprefix  [0]{URL }%
\providecommand \Eprint [0]{\href }%
\providecommand \doibase [0]{https://doi.org/}%
\providecommand \selectlanguage [0]{\@gobble}%
\providecommand \bibinfo  [0]{\@secondoftwo}%
\providecommand \bibfield  [0]{\@secondoftwo}%
\providecommand \translation [1]{[#1]}%
\providecommand \BibitemOpen [0]{}%
\providecommand \bibitemStop [0]{}%
\providecommand \bibitemNoStop [0]{.\EOS\space}%
\providecommand \EOS [0]{\spacefactor3000\relax}%
\providecommand \BibitemShut  [1]{\csname bibitem#1\endcsname}%
\let\auto@bib@innerbib\@empty
\bibitem [{\citenamefont {Gildemeister}\ \emph {et~al.}(1999)\citenamefont
  {Gildemeister}, \citenamefont {Lee},\ and\ \citenamefont
  {Richards}}]{intro1}%
  \BibitemOpen
  \bibfield  {author} {\bibinfo {author} {\bibfnamefont {J.}~\bibnamefont
  {Gildemeister}}, \bibinfo {author} {\bibfnamefont {A.~T.}\ \bibnamefont
  {Lee}},\ and\ \bibinfo {author} {\bibfnamefont {P.}~\bibnamefont
  {Richards}},\ }\bibfield  {title} {``\bibinfo {title} {A fully lithographed
  voltage-biased superconducting spiderweb bolometer}'',\ }\href@noop {}
  {\bibfield  {journal} {\bibinfo  {journal} {Applied Physics Letters}\
  }\textbf {\bibinfo {volume} {74}},\ \bibinfo {pages} {868} (\bibinfo {year}
  {1999})}\BibitemShut {NoStop}%
\bibitem [{\citenamefont {Irwin}(1995)}]{intro2}%
  \BibitemOpen
  \bibfield  {author} {\bibinfo {author} {\bibfnamefont {K.}~\bibnamefont
  {Irwin}},\ }\bibfield  {title} {``\bibinfo {title} {An application of
  electrothermal feedback for high resolution cryogenic particle detection}'',\
  }\href@noop {} {\bibfield  {journal} {\bibinfo  {journal} {Applied Physics
  Letters}\ }\textbf {\bibinfo {volume} {66}},\ \bibinfo {pages} {1998}
  (\bibinfo {year} {1995})}\BibitemShut {NoStop}%
\bibitem [{\citenamefont {Cabrera}\ \emph {et~al.}(1998)\citenamefont
  {Cabrera}, \citenamefont {Clarke}, \citenamefont {Colling}, \citenamefont
  {Miller}, \citenamefont {Nam},\ and\ \citenamefont {Romani}}]{intro3}%
  \BibitemOpen
  \bibfield  {author} {\bibinfo {author} {\bibfnamefont {B.}~\bibnamefont
  {Cabrera}}, \bibinfo {author} {\bibfnamefont {R.}~\bibnamefont {Clarke}},
  \bibinfo {author} {\bibfnamefont {P.}~\bibnamefont {Colling}}, \bibinfo
  {author} {\bibfnamefont {A.}~\bibnamefont {Miller}}, \bibinfo {author}
  {\bibfnamefont {S.}~\bibnamefont {Nam}},\ and\ \bibinfo {author}
  {\bibfnamefont {R.}~\bibnamefont {Romani}},\ }\bibfield  {title} {``\bibinfo
  {title} {Detection of single infrared, optical, and ultraviolet photons using
  superconducting transition edge sensors}'',\ }\href@noop {} {\bibfield
  {journal} {\bibinfo  {journal} {Applied Physics Letters}\ }\textbf {\bibinfo
  {volume} {73}},\ \bibinfo {pages} {735} (\bibinfo {year} {1998})}\BibitemShut
  {NoStop}%
\bibitem [{\citenamefont {Chow}\ \emph {et~al.}(2000)\citenamefont {Chow},
  \citenamefont {Loshak}, \citenamefont {van~den Berg}, \citenamefont {Frank},
  \citenamefont {Barbee~Jr},\ and\ \citenamefont {Labov}}]{intro4}%
  \BibitemOpen
  \bibfield  {author} {\bibinfo {author} {\bibfnamefont {D.~T.}\ \bibnamefont
  {Chow}}, \bibinfo {author} {\bibfnamefont {A.}~\bibnamefont {Loshak}},
  \bibinfo {author} {\bibfnamefont {M.~L.}\ \bibnamefont {van~den Berg}},
  \bibinfo {author} {\bibfnamefont {M.~A.}\ \bibnamefont {Frank}}, \bibinfo
  {author} {\bibfnamefont {T.~W.}\ \bibnamefont {Barbee~Jr}},\ and\ \bibinfo
  {author} {\bibfnamefont {S.~E.}\ \bibnamefont {Labov}},\ }\bibfield  {title}
  {``\bibinfo {title} {High-resolution gamma-ray spectrometers using bulk
  absorbers coupled to mo/cu multilayer superconducting transition-edge
  sensors}'',\ }in\ \href@noop {} {\emph {\bibinfo {booktitle} {Hard X-Ray,
  Gamma-Ray, and Neutron Detector Physics II}}},\ Vol.\ \bibinfo {volume}
  {4141}\ (\bibinfo {organization} {SPIE},\ \bibinfo {year} {2000})\ pp.\
  \bibinfo {pages} {67--75}\BibitemShut {NoStop}%
\bibitem [{\citenamefont {Angloher}\ \emph {et~al.}(2016)\citenamefont
  {Angloher}, \citenamefont {Bauer}, \citenamefont {Ferreiro}, \citenamefont
  {Hauff}, \citenamefont {Tanzke}, \citenamefont {Strauss}, \citenamefont
  {Kiefer}, \citenamefont {Petricia}, \citenamefont {Reindl}, \citenamefont
  {Seidel} \emph {et~al.}}]{cresst}%
  \BibitemOpen
  \bibfield  {author} {\bibinfo {author} {\bibfnamefont {G.}~\bibnamefont
  {Angloher}}, \bibinfo {author} {\bibfnamefont {P.}~\bibnamefont {Bauer}},
  \bibinfo {author} {\bibfnamefont {N.}~\bibnamefont {Ferreiro}}, \bibinfo
  {author} {\bibfnamefont {D.}~\bibnamefont {Hauff}}, \bibinfo {author}
  {\bibfnamefont {A.}~\bibnamefont {Tanzke}}, \bibinfo {author} {\bibfnamefont
  {R.}~\bibnamefont {Strauss}}, \bibinfo {author} {\bibfnamefont
  {M.}~\bibnamefont {Kiefer}}, \bibinfo {author} {\bibfnamefont
  {F.}~\bibnamefont {Petricia}}, \bibinfo {author} {\bibfnamefont
  {F.}~\bibnamefont {Reindl}}, \bibinfo {author} {\bibfnamefont
  {W.}~\bibnamefont {Seidel}}, \emph {et~al.},\ }\bibfield  {title} {``\bibinfo
  {title} {Quasiparticle diffusion in cresst light detectors}'',\ }\href@noop
  {} {\bibfield  {journal} {\bibinfo  {journal} {Journal of Low Temperature
  Physics}\ }\textbf {\bibinfo {volume} {184}},\ \bibinfo {pages} {323}
  (\bibinfo {year} {2016})}\BibitemShut {NoStop}%
\bibitem [{\citenamefont {Saab}\ \emph {et~al.}(2000)\citenamefont {Saab},
  \citenamefont {Clarke}, \citenamefont {Cabrera}, \citenamefont {Abusaidi},\
  and\ \citenamefont {Gaitskell}}]{cdms}%
  \BibitemOpen
  \bibfield  {author} {\bibinfo {author} {\bibfnamefont {T.}~\bibnamefont
  {Saab}}, \bibinfo {author} {\bibfnamefont {R.}~\bibnamefont {Clarke}},
  \bibinfo {author} {\bibfnamefont {B.}~\bibnamefont {Cabrera}}, \bibinfo
  {author} {\bibfnamefont {R.}~\bibnamefont {Abusaidi}},\ and\ \bibinfo
  {author} {\bibfnamefont {R.}~\bibnamefont {Gaitskell}},\ }\bibfield  {title}
  {``\bibinfo {title} {Design of qet phonon sensors for the cdms zip
  detectors}'',\ }\href@noop {} {\bibfield  {journal} {\bibinfo  {journal}
  {Nuclear Instruments and Methods in Physics Research Section A: Accelerators,
  Spectrometers, Detectors and Associated Equipment}\ }\textbf {\bibinfo
  {volume} {444}},\ \bibinfo {pages} {300} (\bibinfo {year}
  {2000})}\BibitemShut {NoStop}%
\bibitem [{\citenamefont {Orlando}\ \emph {et~al.}(2018)\citenamefont
  {Orlando}, \citenamefont {Ceriale}, \citenamefont {Ceruti}, \citenamefont
  {De~Gerone}, \citenamefont {Faverzani}, \citenamefont {Ferri}, \citenamefont
  {Gallucci}, \citenamefont {Giachero}, \citenamefont {Nucciotti},
  \citenamefont {Puiu} \emph {et~al.}}]{holmes}%
  \BibitemOpen
  \bibfield  {author} {\bibinfo {author} {\bibfnamefont {A.}~\bibnamefont
  {Orlando}}, \bibinfo {author} {\bibfnamefont {V.}~\bibnamefont {Ceriale}},
  \bibinfo {author} {\bibfnamefont {G.}~\bibnamefont {Ceruti}}, \bibinfo
  {author} {\bibfnamefont {M.}~\bibnamefont {De~Gerone}}, \bibinfo {author}
  {\bibfnamefont {M.}~\bibnamefont {Faverzani}}, \bibinfo {author}
  {\bibfnamefont {E.}~\bibnamefont {Ferri}}, \bibinfo {author} {\bibfnamefont
  {G.}~\bibnamefont {Gallucci}}, \bibinfo {author} {\bibfnamefont
  {A.}~\bibnamefont {Giachero}}, \bibinfo {author} {\bibfnamefont
  {A.}~\bibnamefont {Nucciotti}}, \bibinfo {author} {\bibfnamefont
  {A.}~\bibnamefont {Puiu}}, \emph {et~al.},\ }\bibfield  {title} {``\bibinfo
  {title} {Microfabrication of transition-edge sensor arrays of
  microcalorimeters with 163ho for direct neutrino mass measurements with
  holmes}'',\ }\href@noop {} {\bibfield  {journal} {\bibinfo  {journal}
  {Journal of Low Temperature Physics}\ }\textbf {\bibinfo {volume} {193}},\
  \bibinfo {pages} {771} (\bibinfo {year} {2018})}\BibitemShut {NoStop}%
\bibitem [{\citenamefont {Irwin}\ and\ \citenamefont {Hilton}(2005)}]{irwin}%
  \BibitemOpen
  \bibfield  {author} {\bibinfo {author} {\bibfnamefont {K.~D.}\ \bibnamefont
  {Irwin}}\ and\ \bibinfo {author} {\bibfnamefont {G.~C.}\ \bibnamefont
  {Hilton}},\ }\bibfield  {title} {``\bibinfo {title} {Transition-edge
  sensors}'',\ }in\ \href@noop {} {\emph {\bibinfo {booktitle} {Cryogenic
  particle detection}}}\ (\bibinfo  {publisher} {Springer},\ \bibinfo {year}
  {2005})\ pp.\ \bibinfo {pages} {63--150}\BibitemShut {NoStop}%
\bibitem [{\citenamefont {Lolli}\ \emph {et~al.}(2013)\citenamefont {Lolli},
  \citenamefont {Taralli}, \citenamefont {Portesi}, \citenamefont {Monticone},\
  and\ \citenamefont {Rajteri}}]{fotonires}%
  \BibitemOpen
  \bibfield  {author} {\bibinfo {author} {\bibfnamefont {L.}~\bibnamefont
  {Lolli}}, \bibinfo {author} {\bibfnamefont {E.}~\bibnamefont {Taralli}},
  \bibinfo {author} {\bibfnamefont {C.}~\bibnamefont {Portesi}}, \bibinfo
  {author} {\bibfnamefont {E.}~\bibnamefont {Monticone}},\ and\ \bibinfo
  {author} {\bibfnamefont {M.}~\bibnamefont {Rajteri}},\ }\bibfield  {title}
  {``\bibinfo {title} {High intrinsic energy resolution photon number resolving
  detectors}'',\ }\href@noop {} {\bibfield  {journal} {\bibinfo  {journal}
  {Applied Physics Letters}\ }\textbf {\bibinfo {volume} {103}},\ \bibinfo
  {pages} {041107} (\bibinfo {year} {2013})}\BibitemShut {NoStop}%
\bibitem [{\citenamefont {Hattori}\ \emph {et~al.}(2022)\citenamefont
  {Hattori}, \citenamefont {Konno}, \citenamefont {Miura}, \citenamefont
  {Takasu},\ and\ \citenamefont {Fukuda}}]{fotonires2}%
  \BibitemOpen
  \bibfield  {author} {\bibinfo {author} {\bibfnamefont {K.}~\bibnamefont
  {Hattori}}, \bibinfo {author} {\bibfnamefont {T.}~\bibnamefont {Konno}},
  \bibinfo {author} {\bibfnamefont {Y.}~\bibnamefont {Miura}}, \bibinfo
  {author} {\bibfnamefont {S.}~\bibnamefont {Takasu}},\ and\ \bibinfo {author}
  {\bibfnamefont {D.}~\bibnamefont {Fukuda}},\ }\bibfield  {title} {``\bibinfo
  {title} {An optical transition-edge sensor with high energy resolution}'',\
  }\href@noop {} {\bibfield  {journal} {\bibinfo  {journal} {Superconductor
  Science and Technology}\ }\textbf {\bibinfo {volume} {35}},\ \bibinfo {pages}
  {095002} (\bibinfo {year} {2022})}\BibitemShut {NoStop}%
\bibitem [{\citenamefont {Apponi}\ \emph {et~al.}(2021)\citenamefont {Apponi},
  \citenamefont {Pandolfi}, \citenamefont {Rago}, \citenamefont {Cavoto},
  \citenamefont {Mariani},\ and\ \citenamefont {Ruocco}}]{mcp1}%
  \BibitemOpen
  \bibfield  {author} {\bibinfo {author} {\bibfnamefont {A.}~\bibnamefont
  {Apponi}}, \bibinfo {author} {\bibfnamefont {F.}~\bibnamefont {Pandolfi}},
  \bibinfo {author} {\bibfnamefont {I.}~\bibnamefont {Rago}}, \bibinfo {author}
  {\bibfnamefont {G.}~\bibnamefont {Cavoto}}, \bibinfo {author} {\bibfnamefont
  {C.}~\bibnamefont {Mariani}},\ and\ \bibinfo {author} {\bibfnamefont
  {A.}~\bibnamefont {Ruocco}},\ }\bibfield  {title} {``\bibinfo {title}
  {Absolute efficiency of a two-stage microchannel plate for electrons in the
  30--900 ev energy range}'',\ }\href@noop {} {\bibfield  {journal} {\bibinfo
  {journal} {Measurement Science and Technology}\ }\textbf {\bibinfo {volume}
  {33}},\ \bibinfo {pages} {025102} (\bibinfo {year} {2021})}\BibitemShut
  {NoStop}%
\bibitem [{\citenamefont {Apponi}\ \emph {et~al.}(2020)\citenamefont {Apponi},
  \citenamefont {Cavoto}, \citenamefont {Iannone}, \citenamefont {Mariani},
  \citenamefont {Pandolfi}, \citenamefont {Paoloni}, \citenamefont {Rago},\
  and\ \citenamefont {Ruocco}}]{mcp2}%
  \BibitemOpen
  \bibfield  {author} {\bibinfo {author} {\bibfnamefont {A.}~\bibnamefont
  {Apponi}}, \bibinfo {author} {\bibfnamefont {G.}~\bibnamefont {Cavoto}},
  \bibinfo {author} {\bibfnamefont {M.}~\bibnamefont {Iannone}}, \bibinfo
  {author} {\bibfnamefont {C.}~\bibnamefont {Mariani}}, \bibinfo {author}
  {\bibfnamefont {F.}~\bibnamefont {Pandolfi}}, \bibinfo {author}
  {\bibfnamefont {D.}~\bibnamefont {Paoloni}}, \bibinfo {author} {\bibfnamefont
  {I.}~\bibnamefont {Rago}},\ and\ \bibinfo {author} {\bibfnamefont
  {A.}~\bibnamefont {Ruocco}},\ }\bibfield  {title} {``\bibinfo {title}
  {Response of windowless silicon avalanche photo-diodes to electrons in the
  90--900 ev range}'',\ }\href@noop {} {\bibfield  {journal} {\bibinfo
  {journal} {Journal of Instrumentation}\ }\textbf {\bibinfo {volume}
  {15}}\bibinfo  {number} { (11)},\ \bibinfo {pages} {P11015}}\BibitemShut
  {NoStop}%
\bibitem [{\citenamefont {Gugiatti}\ \emph {et~al.}(2020)\citenamefont
  {Gugiatti}, \citenamefont {Biassoni}, \citenamefont {Carminati},
  \citenamefont {Cremonesi}, \citenamefont {Fiorini}, \citenamefont {King},
  \citenamefont {Lechner}, \citenamefont {Mertens}, \citenamefont {Pagnanini},
  \citenamefont {Pavan} \emph {et~al.}}]{sdd}%
  \BibitemOpen
\bibfield  {number} {  }\bibfield  {author} {\bibinfo {author} {\bibfnamefont
  {M.}~\bibnamefont {Gugiatti}}, \bibinfo {author} {\bibfnamefont
  {M.}~\bibnamefont {Biassoni}}, \bibinfo {author} {\bibfnamefont
  {M.}~\bibnamefont {Carminati}}, \bibinfo {author} {\bibfnamefont
  {O.}~\bibnamefont {Cremonesi}}, \bibinfo {author} {\bibfnamefont
  {C.}~\bibnamefont {Fiorini}}, \bibinfo {author} {\bibfnamefont
  {P.}~\bibnamefont {King}}, \bibinfo {author} {\bibfnamefont {P.}~\bibnamefont
  {Lechner}}, \bibinfo {author} {\bibfnamefont {S.}~\bibnamefont {Mertens}},
  \bibinfo {author} {\bibfnamefont {L.}~\bibnamefont {Pagnanini}}, \bibinfo
  {author} {\bibfnamefont {M.}~\bibnamefont {Pavan}}, \emph {et~al.},\
  }\bibfield  {title} {``\bibinfo {title} {Characterisation of a silicon drift
  detector for high-resolution electron spectroscopy}'',\ }\href@noop {}
  {\bibfield  {journal} {\bibinfo  {journal} {Nuclear Instruments and Methods
  in Physics Research Section A: Accelerators, Spectrometers, Detectors and
  Associated Equipment}\ }\textbf {\bibinfo {volume} {979}},\ \bibinfo {pages}
  {164474} (\bibinfo {year} {2020})}\BibitemShut {NoStop}%
\bibitem [{\citenamefont {Manenti}\ \emph {et~al.}(2024)\citenamefont
  {Manenti}, \citenamefont {Pepe}, \citenamefont {Sarnoff}, \citenamefont
  {Ibrayev}, \citenamefont {Oikonomou}, \citenamefont {Knyazev}, \citenamefont
  {Monticone}, \citenamefont {Garrone}, \citenamefont {Alder}, \citenamefont
  {Fawwaz} \emph {et~al.}}]{dark_counts}%
  \BibitemOpen
  \bibfield  {author} {\bibinfo {author} {\bibfnamefont {L.}~\bibnamefont
  {Manenti}}, \bibinfo {author} {\bibfnamefont {C.}~\bibnamefont {Pepe}},
  \bibinfo {author} {\bibfnamefont {I.}~\bibnamefont {Sarnoff}}, \bibinfo
  {author} {\bibfnamefont {T.}~\bibnamefont {Ibrayev}}, \bibinfo {author}
  {\bibfnamefont {P.}~\bibnamefont {Oikonomou}}, \bibinfo {author}
  {\bibfnamefont {A.}~\bibnamefont {Knyazev}}, \bibinfo {author} {\bibfnamefont
  {E.}~\bibnamefont {Monticone}}, \bibinfo {author} {\bibfnamefont
  {H.}~\bibnamefont {Garrone}}, \bibinfo {author} {\bibfnamefont
  {F.}~\bibnamefont {Alder}}, \bibinfo {author} {\bibfnamefont
  {O.}~\bibnamefont {Fawwaz}}, \emph {et~al.},\ }\bibfield  {title} {``\bibinfo
  {title} {Dark counts in optical superconducting transition-edge sensors for
  rare-event searches}'',\ }\href@noop {} {\bibfield  {journal} {\bibinfo
  {journal} {Physical Review Applied}\ }\textbf {\bibinfo {volume} {22}},\
  \bibinfo {pages} {024051} (\bibinfo {year} {2024})}\BibitemShut {NoStop}%
\bibitem [{\citenamefont {Lamas-Linares}\ \emph {et~al.}(2013)\citenamefont
  {Lamas-Linares}, \citenamefont {Calkins}, \citenamefont {Tomlin},
  \citenamefont {Gerrits}, \citenamefont {Lita}, \citenamefont {Beyer},
  \citenamefont {Mirin},\ and\ \citenamefont {Woo~Nam}}]{timeres}%
  \BibitemOpen
  \bibfield  {author} {\bibinfo {author} {\bibfnamefont {A.}~\bibnamefont
  {Lamas-Linares}}, \bibinfo {author} {\bibfnamefont {B.}~\bibnamefont
  {Calkins}}, \bibinfo {author} {\bibfnamefont {N.~A.}\ \bibnamefont {Tomlin}},
  \bibinfo {author} {\bibfnamefont {T.}~\bibnamefont {Gerrits}}, \bibinfo
  {author} {\bibfnamefont {A.~E.}\ \bibnamefont {Lita}}, \bibinfo {author}
  {\bibfnamefont {J.}~\bibnamefont {Beyer}}, \bibinfo {author} {\bibfnamefont
  {R.~P.}\ \bibnamefont {Mirin}},\ and\ \bibinfo {author} {\bibfnamefont
  {S.}~\bibnamefont {Woo~Nam}},\ }\bibfield  {title} {``\bibinfo {title}
  {Nanosecond-scale timing jitter for single photon detection in transition
  edge sensors}'',\ }\href@noop {} {\bibfield  {journal} {\bibinfo  {journal}
  {Applied Physics Letters}\ }\textbf {\bibinfo {volume} {102}} (\bibinfo
  {year} {2013})}\BibitemShut {NoStop}%
\bibitem [{\citenamefont {Kobayashi}\ \emph {et~al.}(2019)\citenamefont
  {Kobayashi}, \citenamefont {Hattori}, \citenamefont {Inoue},\ and\
  \citenamefont {Fukuda}}]{maxcount}%
  \BibitemOpen
  \bibfield  {author} {\bibinfo {author} {\bibfnamefont {R.}~\bibnamefont
  {Kobayashi}}, \bibinfo {author} {\bibfnamefont {K.}~\bibnamefont {Hattori}},
  \bibinfo {author} {\bibfnamefont {S.}~\bibnamefont {Inoue}},\ and\ \bibinfo
  {author} {\bibfnamefont {D.}~\bibnamefont {Fukuda}},\ }\bibfield  {title}
  {``\bibinfo {title} {Development of a fast response titanium-gold bilayer
  optical tes with an optical fiber self-alignment structure}'',\ }\href@noop
  {} {\bibfield  {journal} {\bibinfo  {journal} {IEEE Transactions on Applied
  Superconductivity}\ }\textbf {\bibinfo {volume} {29}},\ \bibinfo {pages} {1}
  (\bibinfo {year} {2019})}\BibitemShut {NoStop}%
\bibitem [{\citenamefont {Patel}\ \emph {et~al.}(2024)\citenamefont {Patel},
  \citenamefont {Withington}, \citenamefont {Shard}, \citenamefont {Goldie},\
  and\ \citenamefont {Thomas}}]{patel}%
  \BibitemOpen
  \bibfield  {author} {\bibinfo {author} {\bibfnamefont {K.}~\bibnamefont
  {Patel}}, \bibinfo {author} {\bibfnamefont {S.}~\bibnamefont {Withington}},
  \bibinfo {author} {\bibfnamefont {A.}~\bibnamefont {Shard}}, \bibinfo
  {author} {\bibfnamefont {D.}~\bibnamefont {Goldie}},\ and\ \bibinfo {author}
  {\bibfnamefont {C.}~\bibnamefont {Thomas}},\ }\bibfield  {title} {``\bibinfo
  {title} {Electron spectroscopy using transition-edge sensors}'',\ }\href@noop
  {} {\bibfield  {journal} {\bibinfo  {journal} {Journal of Applied Physics}\
  }\textbf {\bibinfo {volume} {135}},\ \bibinfo {pages} {224504} (\bibinfo
  {year} {2024})}\BibitemShut {NoStop}%
\bibitem [{\citenamefont {Loidl}\ \emph {et~al.}(2019)\citenamefont {Loidl},
  \citenamefont {Beyer}, \citenamefont {Bockhorn}, \citenamefont {Enss},
  \citenamefont {Kempf}, \citenamefont {Kossert}, \citenamefont {Mariam},
  \citenamefont {N{\"a}hle}, \citenamefont {Paulsen}, \citenamefont {Ranitzsch}
  \emph {et~al.}}]{microcal1}%
  \BibitemOpen
  \bibfield  {author} {\bibinfo {author} {\bibfnamefont {M.}~\bibnamefont
  {Loidl}}, \bibinfo {author} {\bibfnamefont {J.}~\bibnamefont {Beyer}},
  \bibinfo {author} {\bibfnamefont {L.}~\bibnamefont {Bockhorn}}, \bibinfo
  {author} {\bibfnamefont {C.}~\bibnamefont {Enss}}, \bibinfo {author}
  {\bibfnamefont {S.}~\bibnamefont {Kempf}}, \bibinfo {author} {\bibfnamefont
  {K.}~\bibnamefont {Kossert}}, \bibinfo {author} {\bibfnamefont
  {R.}~\bibnamefont {Mariam}}, \bibinfo {author} {\bibfnamefont
  {O.}~\bibnamefont {N{\"a}hle}}, \bibinfo {author} {\bibfnamefont
  {M.}~\bibnamefont {Paulsen}}, \bibinfo {author} {\bibfnamefont
  {P.}~\bibnamefont {Ranitzsch}}, \emph {et~al.},\ }\bibfield  {title}
  {``\bibinfo {title} {Beta spectrometry with metallic magnetic calorimeters in
  the framework of the european empir project metrobeta}'',\ }\href@noop {}
  {\bibfield  {journal} {\bibinfo  {journal} {Applied Radiation and Isotopes}\
  }\textbf {\bibinfo {volume} {153}},\ \bibinfo {pages} {108830} (\bibinfo
  {year} {2019})}\BibitemShut {NoStop}%
\bibitem [{\citenamefont {Fitzgerald}\ \emph {et~al.}(2022)\citenamefont
  {Fitzgerald}, \citenamefont {Alpert}, \citenamefont {Becker}, \citenamefont
  {Bergeron}, \citenamefont {Essex}, \citenamefont {Morgan}, \citenamefont
  {Nour}, \citenamefont {O’Neil}, \citenamefont {Schmidt}, \citenamefont
  {Shaw} \emph {et~al.}}]{microcal2}%
  \BibitemOpen
  \bibfield  {author} {\bibinfo {author} {\bibfnamefont {R.~P.}\ \bibnamefont
  {Fitzgerald}}, \bibinfo {author} {\bibfnamefont {B.~K.}\ \bibnamefont
  {Alpert}}, \bibinfo {author} {\bibfnamefont {D.~T.}\ \bibnamefont {Becker}},
  \bibinfo {author} {\bibfnamefont {D.~E.}\ \bibnamefont {Bergeron}}, \bibinfo
  {author} {\bibfnamefont {R.~M.}\ \bibnamefont {Essex}}, \bibinfo {author}
  {\bibfnamefont {K.}~\bibnamefont {Morgan}}, \bibinfo {author} {\bibfnamefont
  {S.}~\bibnamefont {Nour}}, \bibinfo {author} {\bibfnamefont {G.}~\bibnamefont
  {O’Neil}}, \bibinfo {author} {\bibfnamefont {D.~R.}\ \bibnamefont
  {Schmidt}}, \bibinfo {author} {\bibfnamefont {G.~A.}\ \bibnamefont {Shaw}},
  \emph {et~al.},\ }\bibfield  {title} {``\bibinfo {title} {Toward a new
  primary standardization of radionuclide massic activity using
  microcalorimetry and quantitative milligram-scale samples}'',\ }\href@noop {}
  {\bibfield  {journal} {\bibinfo  {journal} {Journal of Research of the
  National Institute of Standards and Technology}\ }\textbf {\bibinfo {volume}
  {126}},\ \bibinfo {pages} {126048} (\bibinfo {year} {2022})}\BibitemShut
  {NoStop}%
\bibitem [{\citenamefont {Betti}\ \emph
  {et~al.}(2019{\natexlab{a}})\citenamefont {Betti}, \citenamefont {Biasotti},
  \citenamefont {Bosc{\'a}}, \citenamefont {Calle}, \citenamefont
  {Carabe-Lopez}, \citenamefont {Cavoto}, \citenamefont {Chang}, \citenamefont
  {Chung}, \citenamefont {Cocco}, \citenamefont {Colijn} \emph
  {et~al.}}]{ptolemy1}%
  \BibitemOpen
  \bibfield  {author} {\bibinfo {author} {\bibfnamefont {M.}~\bibnamefont
  {Betti}}, \bibinfo {author} {\bibfnamefont {M.}~\bibnamefont {Biasotti}},
  \bibinfo {author} {\bibfnamefont {A.}~\bibnamefont {Bosc{\'a}}}, \bibinfo
  {author} {\bibfnamefont {F.}~\bibnamefont {Calle}}, \bibinfo {author}
  {\bibfnamefont {J.}~\bibnamefont {Carabe-Lopez}}, \bibinfo {author}
  {\bibfnamefont {G.}~\bibnamefont {Cavoto}}, \bibinfo {author} {\bibfnamefont
  {C.}~\bibnamefont {Chang}}, \bibinfo {author} {\bibfnamefont
  {W.}~\bibnamefont {Chung}}, \bibinfo {author} {\bibfnamefont
  {A.}~\bibnamefont {Cocco}}, \bibinfo {author} {\bibfnamefont
  {A.}~\bibnamefont {Colijn}}, \emph {et~al.},\ }\bibfield  {title} {``\bibinfo
  {title} {A design for an electromagnetic filter for precision energy
  measurements at the tritium endpoint}'',\ }\href@noop {} {\bibfield
  {journal} {\bibinfo  {journal} {Progress in Particle and Nuclear Physics}\
  }\textbf {\bibinfo {volume} {106}},\ \bibinfo {pages} {120} (\bibinfo {year}
  {2019}{\natexlab{a}})}\BibitemShut {NoStop}%
\bibitem [{\citenamefont {Betti}\ \emph
  {et~al.}(2019{\natexlab{b}})\citenamefont {Betti}, \citenamefont {Biasotti},
  \citenamefont {Bosc{\'a}}, \citenamefont {Calle}, \citenamefont {Canci},
  \citenamefont {Cavoto}, \citenamefont {Chang}, \citenamefont {Cocco},
  \citenamefont {Colijn}, \citenamefont {Conrad} \emph {et~al.}}]{ptolemy2}%
  \BibitemOpen
  \bibfield  {author} {\bibinfo {author} {\bibfnamefont {M.}~\bibnamefont
  {Betti}}, \bibinfo {author} {\bibfnamefont {M.}~\bibnamefont {Biasotti}},
  \bibinfo {author} {\bibfnamefont {A.}~\bibnamefont {Bosc{\'a}}}, \bibinfo
  {author} {\bibfnamefont {F.}~\bibnamefont {Calle}}, \bibinfo {author}
  {\bibfnamefont {N.}~\bibnamefont {Canci}}, \bibinfo {author} {\bibfnamefont
  {G.}~\bibnamefont {Cavoto}}, \bibinfo {author} {\bibfnamefont
  {C.}~\bibnamefont {Chang}}, \bibinfo {author} {\bibfnamefont
  {A.}~\bibnamefont {Cocco}}, \bibinfo {author} {\bibfnamefont
  {A.}~\bibnamefont {Colijn}}, \bibinfo {author} {\bibfnamefont
  {J.}~\bibnamefont {Conrad}}, \emph {et~al.},\ }\bibfield  {title} {``\bibinfo
  {title} {Neutrino physics with the ptolemy project: active neutrino
  properties and the light sterile case}'',\ }\href@noop {} {\bibfield
  {journal} {\bibinfo  {journal} {Journal of Cosmology and Astroparticle
  Physics}\ }\textbf {\bibinfo {volume} {2019}}\bibinfo  {number} { (07)},\
  \bibinfo {pages} {047}}\BibitemShut {NoStop}%
\bibitem [{\citenamefont {Apponi}\ \emph {et~al.}(2022)\citenamefont {Apponi},
  \citenamefont {Betti}, \citenamefont {Borghesi}, \citenamefont {Canci},
  \citenamefont {Cavoto}, \citenamefont {Chang}, \citenamefont {Chung},
  \citenamefont {Cocco}, \citenamefont {Colijn}, \citenamefont {D'Ambrosio}
  \emph {et~al.}}]{ptolemy3}%
  \BibitemOpen
\bibfield  {number} {  }\bibfield  {author} {\bibinfo {author} {\bibfnamefont
  {A.}~\bibnamefont {Apponi}}, \bibinfo {author} {\bibfnamefont
  {M.}~\bibnamefont {Betti}}, \bibinfo {author} {\bibfnamefont
  {M.}~\bibnamefont {Borghesi}}, \bibinfo {author} {\bibfnamefont
  {N.}~\bibnamefont {Canci}}, \bibinfo {author} {\bibfnamefont
  {G.}~\bibnamefont {Cavoto}}, \bibinfo {author} {\bibfnamefont
  {C.}~\bibnamefont {Chang}}, \bibinfo {author} {\bibfnamefont
  {W.}~\bibnamefont {Chung}}, \bibinfo {author} {\bibfnamefont
  {A.}~\bibnamefont {Cocco}}, \bibinfo {author} {\bibfnamefont
  {A.}~\bibnamefont {Colijn}}, \bibinfo {author} {\bibfnamefont
  {N.}~\bibnamefont {D'Ambrosio}}, \emph {et~al.},\ }\bibfield  {title}
  {``\bibinfo {title} {Implementation and optimization of the ptolemy
  transverse drift electromagnetic filter}'',\ }\href@noop {} {\bibfield
  {journal} {\bibinfo  {journal} {Journal of Instrumentation}\ }\textbf
  {\bibinfo {volume} {17}}\bibinfo  {number} { (05)},\ \bibinfo {pages}
  {P05021}}\BibitemShut {NoStop}%
\bibitem [{\citenamefont {Pepe}\ \emph {et~al.}(2024)\citenamefont {Pepe},
  \citenamefont {Corcione}, \citenamefont {Pandolfi}, \citenamefont {Garrone},
  \citenamefont {Monticone}, \citenamefont {Rago}, \citenamefont {Cavoto},
  \citenamefont {Apponi}, \citenamefont {Ruocco}, \citenamefont {Malnati} \emph
  {et~al.}}]{articolo}%
  \BibitemOpen
\bibfield  {number} {  }\bibfield  {author} {\bibinfo {author} {\bibfnamefont
  {C.}~\bibnamefont {Pepe}}, \bibinfo {author} {\bibfnamefont {B.}~\bibnamefont
  {Corcione}}, \bibinfo {author} {\bibfnamefont {F.}~\bibnamefont {Pandolfi}},
  \bibinfo {author} {\bibfnamefont {H.}~\bibnamefont {Garrone}}, \bibinfo
  {author} {\bibfnamefont {E.}~\bibnamefont {Monticone}}, \bibinfo {author}
  {\bibfnamefont {I.}~\bibnamefont {Rago}}, \bibinfo {author} {\bibfnamefont
  {G.}~\bibnamefont {Cavoto}}, \bibinfo {author} {\bibfnamefont
  {A.}~\bibnamefont {Apponi}}, \bibinfo {author} {\bibfnamefont
  {A.}~\bibnamefont {Ruocco}}, \bibinfo {author} {\bibfnamefont
  {F.}~\bibnamefont {Malnati}}, \emph {et~al.},\ }\bibfield  {title}
  {``\bibinfo {title} {Detection of low-energy electrons with transition-edge
  sensors}'',\ }\href@noop {} {\bibfield  {journal} {\bibinfo  {journal}
  {Physical Review Applied}\ }\textbf {\bibinfo {volume} {22}},\ \bibinfo
  {pages} {L041007} (\bibinfo {year} {2024})}\BibitemShut {NoStop}%
\bibitem [{\citenamefont {Corcione}(2025)}]{ncc}%
  \BibitemOpen
  \bibfield  {author} {\bibinfo {author} {\bibfnamefont {B.}~\bibnamefont
  {Corcione}},\ }\bibfield  {title} {``\bibinfo {title} {Detection of
  low-energy electrons with transition-edge sensors}'',\ }\href
  {https://doi.org/10.1393/ncc/i2025-25187-9} {\bibfield  {journal} {\bibinfo
  {journal} {Il Nuovo Cimento C}\ }\textbf {\bibinfo {volume} {48}},\ \bibinfo
  {pages} {187} (\bibinfo {year} {2025})},\ \bibinfo {note} {published online
  26 August 2025}\BibitemShut {NoStop}%
\bibitem [{\citenamefont {Drung}\ \emph {et~al.}(2007)\citenamefont {Drung},
  \citenamefont {Abmann}, \citenamefont {Beyer}, \citenamefont {Kirste},
  \citenamefont {Peters}, \citenamefont {Ruede},\ and\ \citenamefont
  {Schurig}}]{squid}%
  \BibitemOpen
  \bibfield  {author} {\bibinfo {author} {\bibfnamefont {D.}~\bibnamefont
  {Drung}}, \bibinfo {author} {\bibfnamefont {C.}~\bibnamefont {Abmann}},
  \bibinfo {author} {\bibfnamefont {J.}~\bibnamefont {Beyer}}, \bibinfo
  {author} {\bibfnamefont {A.}~\bibnamefont {Kirste}}, \bibinfo {author}
  {\bibfnamefont {M.}~\bibnamefont {Peters}}, \bibinfo {author} {\bibfnamefont
  {F.}~\bibnamefont {Ruede}},\ and\ \bibinfo {author} {\bibfnamefont
  {T.}~\bibnamefont {Schurig}},\ }\bibfield  {title} {``\bibinfo {title}
  {Highly sensitive and easy-to-use squid sensors}'',\ }\href@noop {}
  {\bibfield  {journal} {\bibinfo  {journal} {IEEE Transactions on Applied
  Superconductivity}\ }\textbf {\bibinfo {volume} {17}},\ \bibinfo {pages}
  {699} (\bibinfo {year} {2007})}\BibitemShut {NoStop}%
\bibitem [{\citenamefont {Cecchini}\ \emph {et~al.}(2025)\citenamefont
  {Cecchini}, \citenamefont {Pepe}, \citenamefont {Corcione}, \citenamefont
  {Castellano}, \citenamefont {Paoloni}, \citenamefont {Malnati}, \citenamefont
  {Cavoto}, \citenamefont {Carminati}, \citenamefont {Fiorini}, \citenamefont
  {Pettinari} \emph {et~al.}}]{articolo_cnt}%
  \BibitemOpen
  \bibfield  {author} {\bibinfo {author} {\bibfnamefont {L.}~\bibnamefont
  {Cecchini}}, \bibinfo {author} {\bibfnamefont {C.}~\bibnamefont {Pepe}},
  \bibinfo {author} {\bibfnamefont {B.}~\bibnamefont {Corcione}}, \bibinfo
  {author} {\bibfnamefont {O.}~\bibnamefont {Castellano}}, \bibinfo {author}
  {\bibfnamefont {D.}~\bibnamefont {Paoloni}}, \bibinfo {author} {\bibfnamefont
  {F.}~\bibnamefont {Malnati}}, \bibinfo {author} {\bibfnamefont
  {G.}~\bibnamefont {Cavoto}}, \bibinfo {author} {\bibfnamefont
  {M.}~\bibnamefont {Carminati}}, \bibinfo {author} {\bibfnamefont
  {C.}~\bibnamefont {Fiorini}}, \bibinfo {author} {\bibfnamefont
  {G.}~\bibnamefont {Pettinari}}, \emph {et~al.},\ }\bibfield  {title}
  {``\bibinfo {title} {Quantitative correlation between carbon nanotube tip
  morphology and field emission properties at cryogenic temperature}'',\
  }\href@noop {} {\bibfield  {journal} {\bibinfo  {journal} {Nanoscale}\
  }\textbf {\bibinfo {volume} {17}},\ \bibinfo {pages} {21260} (\bibinfo {year}
  {2025})}\BibitemShut {NoStop}%
\bibitem [{\citenamefont {Schifano}\ \emph {et~al.}(2023)\citenamefont
  {Schifano}, \citenamefont {Cavoto}, \citenamefont {Pandolfi}, \citenamefont
  {Pettinari}, \citenamefont {Apponi}, \citenamefont {Ruocco}, \citenamefont
  {Uccelletti},\ and\ \citenamefont {Rago}}]{titan1}%
  \BibitemOpen
  \bibfield  {author} {\bibinfo {author} {\bibfnamefont {E.}~\bibnamefont
  {Schifano}}, \bibinfo {author} {\bibfnamefont {G.}~\bibnamefont {Cavoto}},
  \bibinfo {author} {\bibfnamefont {F.}~\bibnamefont {Pandolfi}}, \bibinfo
  {author} {\bibfnamefont {G.}~\bibnamefont {Pettinari}}, \bibinfo {author}
  {\bibfnamefont {A.}~\bibnamefont {Apponi}}, \bibinfo {author} {\bibfnamefont
  {A.}~\bibnamefont {Ruocco}}, \bibinfo {author} {\bibfnamefont
  {D.}~\bibnamefont {Uccelletti}},\ and\ \bibinfo {author} {\bibfnamefont
  {I.}~\bibnamefont {Rago}},\ }\bibfield  {title} {``\bibinfo {title}
  {Plasma-etched vertically aligned cnts with enhanced antibacterial power}'',\
  }\href@noop {} {\bibfield  {journal} {\bibinfo  {journal} {Nanomaterials}\
  }\textbf {\bibinfo {volume} {13}},\ \bibinfo {pages} {1081} (\bibinfo {year}
  {2023})}\BibitemShut {NoStop}%
\bibitem [{\citenamefont {Yadav}\ \emph {et~al.}(2024)\citenamefont {Yadav},
  \citenamefont {Rago}, \citenamefont {Pandolfi}, \citenamefont {Mariani},
  \citenamefont {Ruocco}, \citenamefont {Tayyab}, \citenamefont {Apponi},\ and\
  \citenamefont {Cavoto}}]{titan2}%
  \BibitemOpen
  \bibfield  {author} {\bibinfo {author} {\bibfnamefont {R.~P.}\ \bibnamefont
  {Yadav}}, \bibinfo {author} {\bibfnamefont {I.}~\bibnamefont {Rago}},
  \bibinfo {author} {\bibfnamefont {F.}~\bibnamefont {Pandolfi}}, \bibinfo
  {author} {\bibfnamefont {C.}~\bibnamefont {Mariani}}, \bibinfo {author}
  {\bibfnamefont {A.}~\bibnamefont {Ruocco}}, \bibinfo {author} {\bibfnamefont
  {S.}~\bibnamefont {Tayyab}}, \bibinfo {author} {\bibfnamefont
  {A.}~\bibnamefont {Apponi}},\ and\ \bibinfo {author} {\bibfnamefont
  {G.}~\bibnamefont {Cavoto}},\ }\bibfield  {title} {``\bibinfo {title}
  {Evaluation of vertical alignment in carbon nanotubes: A quantitative
  approach}'',\ }\href@noop {} {\bibfield  {journal} {\bibinfo  {journal}
  {Nuclear Instruments and Methods in Physics Research Section A: Accelerators,
  Spectrometers, Detectors and Associated Equipment}\ }\textbf {\bibinfo
  {volume} {1060}},\ \bibinfo {pages} {169081} (\bibinfo {year}
  {2024})}\BibitemShut {NoStop}%
\bibitem [{\citenamefont {Sarasini}\ \emph {et~al.}(2022)\citenamefont
  {Sarasini}, \citenamefont {Tirill{\`o}}, \citenamefont {Lilli}, \citenamefont
  {Bracciale}, \citenamefont {Vullum}, \citenamefont {Berto}, \citenamefont
  {De~Bellis}, \citenamefont {Tamburrano}, \citenamefont {Cavoto},
  \citenamefont {Pandolfi} \emph {et~al.}}]{titan3}%
  \BibitemOpen
  \bibfield  {author} {\bibinfo {author} {\bibfnamefont {F.}~\bibnamefont
  {Sarasini}}, \bibinfo {author} {\bibfnamefont {J.}~\bibnamefont
  {Tirill{\`o}}}, \bibinfo {author} {\bibfnamefont {M.}~\bibnamefont {Lilli}},
  \bibinfo {author} {\bibfnamefont {M.~P.}\ \bibnamefont {Bracciale}}, \bibinfo
  {author} {\bibfnamefont {P.~E.}\ \bibnamefont {Vullum}}, \bibinfo {author}
  {\bibfnamefont {F.}~\bibnamefont {Berto}}, \bibinfo {author} {\bibfnamefont
  {G.}~\bibnamefont {De~Bellis}}, \bibinfo {author} {\bibfnamefont
  {A.}~\bibnamefont {Tamburrano}}, \bibinfo {author} {\bibfnamefont
  {G.}~\bibnamefont {Cavoto}}, \bibinfo {author} {\bibfnamefont
  {F.}~\bibnamefont {Pandolfi}}, \emph {et~al.},\ }\bibfield  {title}
  {``\bibinfo {title} {Highly aligned growth of carbon nanotube forests with
  in-situ catalyst generation: A route to multifunctional basalt fibres}'',\
  }\href@noop {} {\bibfield  {journal} {\bibinfo  {journal} {Composites Part B:
  Engineering}\ }\textbf {\bibinfo {volume} {243}},\ \bibinfo {pages} {110136}
  (\bibinfo {year} {2022})}\BibitemShut {NoStop}%
\bibitem [{\citenamefont {Tayyab}\ \emph {et~al.}(2023)\citenamefont {Tayyab},
  \citenamefont {Apponi}, \citenamefont {Betti}, \citenamefont {Blundo},
  \citenamefont {Cavoto}, \citenamefont {Frisenda}, \citenamefont
  {Jim{\'e}nez-Ar{\'e}valo}, \citenamefont {Mariani}, \citenamefont {Pandolfi},
  \citenamefont {Polimeni} \emph {et~al.}}]{titan4}%
  \BibitemOpen
  \bibfield  {author} {\bibinfo {author} {\bibfnamefont {S.}~\bibnamefont
  {Tayyab}}, \bibinfo {author} {\bibfnamefont {A.}~\bibnamefont {Apponi}},
  \bibinfo {author} {\bibfnamefont {M.~G.}\ \bibnamefont {Betti}}, \bibinfo
  {author} {\bibfnamefont {E.}~\bibnamefont {Blundo}}, \bibinfo {author}
  {\bibfnamefont {G.}~\bibnamefont {Cavoto}}, \bibinfo {author} {\bibfnamefont
  {R.}~\bibnamefont {Frisenda}}, \bibinfo {author} {\bibfnamefont
  {N.}~\bibnamefont {Jim{\'e}nez-Ar{\'e}valo}}, \bibinfo {author}
  {\bibfnamefont {C.}~\bibnamefont {Mariani}}, \bibinfo {author} {\bibfnamefont
  {F.}~\bibnamefont {Pandolfi}}, \bibinfo {author} {\bibfnamefont
  {A.}~\bibnamefont {Polimeni}}, \emph {et~al.},\ }\bibfield  {title}
  {``\bibinfo {title} {Spectromicroscopy study of induced defects in
  ion-bombarded highly aligned carbon nanotubes}'',\ }\href@noop {} {\bibfield
  {journal} {\bibinfo  {journal} {Nanomaterials}\ }\textbf {\bibinfo {volume}
  {14}},\ \bibinfo {pages} {77} (\bibinfo {year} {2023})}\BibitemShut {NoStop}%
\bibitem [{\citenamefont {Fowler}\ and\ \citenamefont
  {Nordheim}(1928)}]{fn_theory}%
  \BibitemOpen
  \bibfield  {author} {\bibinfo {author} {\bibfnamefont {R.~H.}\ \bibnamefont
  {Fowler}}\ and\ \bibinfo {author} {\bibfnamefont {L.}~\bibnamefont
  {Nordheim}},\ }\bibfield  {title} {``\bibinfo {title} {Electron emission in
  intense electric fields}'',\ }\href@noop {} {\bibfield  {journal} {\bibinfo
  {journal} {Proceedings of the royal society of London. Series A, containing
  papers of a mathematical and physical character}\ }\textbf {\bibinfo {volume}
  {119}},\ \bibinfo {pages} {173} (\bibinfo {year} {1928})}\BibitemShut
  {NoStop}%
\bibitem [{\citenamefont {Gomer}(1994)}]{monochromaticity}%
  \BibitemOpen
  \bibfield  {author} {\bibinfo {author} {\bibfnamefont {R.}~\bibnamefont
  {Gomer}},\ }\bibfield  {title} {``\bibinfo {title} {Field emission, field
  ionization, and field desorption}'',\ }\href@noop {} {\bibfield  {journal}
  {\bibinfo  {journal} {Surface science}\ }\textbf {\bibinfo {volume} {299}},\
  \bibinfo {pages} {129} (\bibinfo {year} {1994})}\BibitemShut {NoStop}%
\bibitem [{\citenamefont {Oreglia}(1980)}]{cb}%
  \BibitemOpen
  \bibfield  {author} {\bibinfo {author} {\bibfnamefont {M.~J.}\ \bibnamefont
  {Oreglia}},\ }\href@noop {} {\emph {\bibinfo {title} {Study of the reactions
  $\psi$'→ $\gamma$$\gamma$$\psi$ [Thesis]}}},\ \bibinfo {type} {Tech. Rep.}\
  (\bibinfo  {institution} {SLAC National Accelerator Laboratory (SLAC), Menlo
  Park, CA (United States)},\ \bibinfo {year} {1980})\BibitemShut {NoStop}%
\bibitem [{dat()}]{data_paper_2025}%
  \BibitemOpen
  \href@noop {} {\bibinfo {title} {Shared data for this paper:}}\ \bibinfo
  {howpublished} {\url{https://zenodo.org/records/18786228}}\BibitemShut
  {NoStop}%
\end{thebibliography}%

\end{document}